\documentclass[nologo,11pt,a4paper]{ETHpaper}

 \usepackage{amssymb,amsmath}                   % Mathematical Symbols
 \usepackage{graphicx}                          % Figures
 \usepackage{tikz}
 \usetikzlibrary{arrows,automata,shapes}
 \usepackage[english]{babel}                    % Spell-checking
 \usepackage[justification=centering]{caption}
 \usepackage[]{hyperref} % Hyperlinks

 \usepackage[utf8]{inputenc}

\usepackage{xcolor}

\title{Emotional persistence in online chatting communities}

\author{Antonios Garas$^{1}$, David Garcia$^{1}$, Marcin Skowron$^{2}$
  and Frank Schweitzer$^{1}$}

\address{$^{1}$Chair of Systems Design, ETH Zurich, Kreuzplatz 5, 8032
  Zurich, Switzerland \\ $^{2}$ Austrian Research Institute for
  Artificial Intelligence, Freyung 6/6, 1010 Vienna, Austria}

\date{today}

\authoralternative{A. Garas, D. Garcia, M. Skowron, F. Schweitzer}

\reference{\textsl{Scientific Reports} \textbf{2} 402 (2012)}

\www{\url{http://www.sg.ethz.ch}}

\newcommand{\mean}[1]{\left\langle #1 \right\rangle}

\begin{document}

\maketitle

%%%%%%%%%%%%%%%%%%%%%%%%%%%%%%%%%%%%%%%%%%%%%%%%%%%%%%%%%%%%%%%%

\begin{abstract} 
  How do users behave in online chatrooms, where they instantaneously
  read and write posts? We analyzed about 2.5 million posts covering
  various topics in Internet relay channels, and found that user activity
  patterns follow known power-law and stretched exponential
  distributions, indicating that online chat activity is not different
  from other forms of communication. Analysing the emotional expressions
  (positive, negative, neutral) of users, we revealed a remarkable
  persistence both for individual users and channels. I.e. despite their
  anonymity, users tend to follow social norms in repeated interactions
  in online chats, which results in a specific emotional "tone" of the
  channels. We provide an agent-based model of emotional interaction,
  which recovers qualitatively both the activity patterns in chatrooms
  and the emotional persistence of users and channels. While our
  assumptions about agent's emotional expressions are rooted in
  psychology, the model allows to test different hypothesis regarding
  their emotional impact in online communication.
\end{abstract}

\section*{Introduction}

How do human communication patterns change on the Internet?  Round the
clock activities of Internet users put us into the comfortable situation
of having massive data from various sources available at a fine time
resolution. But what to look at? Which aggregated measures are most
appropriate to capture how new technologies affect our communicative
behavior? And then, are we able to match these findings with a dynamic
model that is able to generate insights into their origin? In this paper,
we provide both: a new way of analysing data from online chats, and a
model of interacting agents to reproduce the stylized facts of our
analysis. In addition to the activity patterns of users, we also analyse
and model their emotional expressions that trigger the interactions of
users in online chats. Validating our agent-based model against empirical
findings allows us to draw conclusions about the role of emotions in this
form of communication.

Online communication can be seen as a large-scale social experiment that
constantly provides us with data about user activities and
interactions. Consequently, time series analyses have already revealed
remarkable temporal activity patterns, e.g. in email communication. Such
patterns allow conclusions how humans organize their time and give
different priorities to their communication tasks
~\cite{Oliveira2005,Barabasi2005,Malmgren2008,Rybski2009,Grinstein2008,Crane2010}.
One particular quantity to describe these patterns is the distribution
$P(\tau)$ of the waiting time $\tau$ that elapses before a particular
user answers e.g. an email. Different studies have confirmed the
power-law nature of this distribution, $P(\tau)\sim \tau ^{-\alpha}$.
Its origin was attributed either to the burstiness of events
\cite{Barabasi2005} or to circadian activity patterns
\cite{Malmgren2008}, while a recent work shows that a combination of both
effects is also a plausible scenario \cite{Jo2012}. However, the value of
the exponent $\alpha$ is still debated. A stochastic priority queue model
\cite{Grinstein2008} allows to derive $\alpha$ by comparing two different
rates, the average rate $\lambda$ of messages arriving and the average
rate $\mu$ of processing messages. If $\mu \leq \lambda$, i.e. if
messages arrive faster than they can be processed, $\alpha=3/2$ was
found, which is compatible with most empirical findings and simulation
models
\cite{Barabasi2005,Vazquez2006,Oliveira2005,Malmgren2008}. However, in
the opposite case, $\mu \geq \lambda$, i.e. if messages can be processed
upon arrival, $\alpha=5/2$ was found together with an exponential
correction term. The latter regime, also denoted as the "highly attentive
regime", could be verified empirically so far only by using data about
donations \cite{Crane2010}.  So, it is an interesting question to analyze
other forms of online communication to see whether there is evidence for
the second regime.

In this paper, we analyze data about instant online communication in
different chatting communities, specifically Internet Relay Chat (IRC)
channels, where each channel covers a particular topic. Prior to the very
common social networking sites of today, IRC channels provided a safe and
independent way for users to share and discuss information outside
traditional media. Different from other types of online communication,
such as blogs or fora where entries are posted at a given time (decided
by the writer), IRC chats are instantaneous in real time, i.e. users read
while the post is written and can react immediately.  This type of
interaction requires much higher user activity in comparison to
persistent communication e.g. in fora. Further, it is more spontaneous,
often leading to emotionally-rich communication between involved
peers. Consequently, instant communication should require specific tools
and models for analysis, that are capable of covering these predominant
features.

Nowadays, IRC channels are still one of the most used platforms for
collective real-time online communication and are used for various
purposes, e.g. organization of open-source project development, Internet
activism, dating, etc. Our dataset (described in detail in the data
section), consists of 20 IRC channels covering topics as diverse as
music, sports, casuals chats, business, politics, or computer related
issues -- which is important to ensure that there is no topical bias
involved in our analysis. For each channel, we have consecutive daily
recordings of the open discussion over a period of 42 days, which amounts
to more than 2.5 million posts in total generated by more than 20.000
different users.

We process our analysis as follows: first, we look into the communication
patterns of instant online discussions, to find out about the average
response time of users and its possible dependence on the topics
discussed. This shall allow us to identify differences between
instantaneous chatting communities and other forms of slower, persistent
communication. In a second step, we look more closely into the content of
the discussions and how they depend on the emotions expressed by users.
Remarkably, we find that most users are very persistent in expressing
their positive or negative emotions - which is not expected given the
variety of topics and the user anonymity. This leads us to the question
in what respect online chats are different from offline discussions which
are mostly guided by social norms. We argue that even in instantaneous,
anonymous online chats users behave very much like "normal" people.  Our
quantitative insights into user's activity patters and their emotional
expressions are eventually combined to model interacting emotional
agents. We demonstrate that the stylised facts of the emotional
persistence can be reproduced by our model by only calibrating a small
set of agent features. This success indicates that our modeling framework
can be used to test further hypothesis about emotional interaction in
online communities.

\section*{Results}
\subsection*{User activity patterns}
\label{sec:userAct}

\begin{figure}[h]
 \centering
% Figure 1
 \includegraphics[scale=0.6]{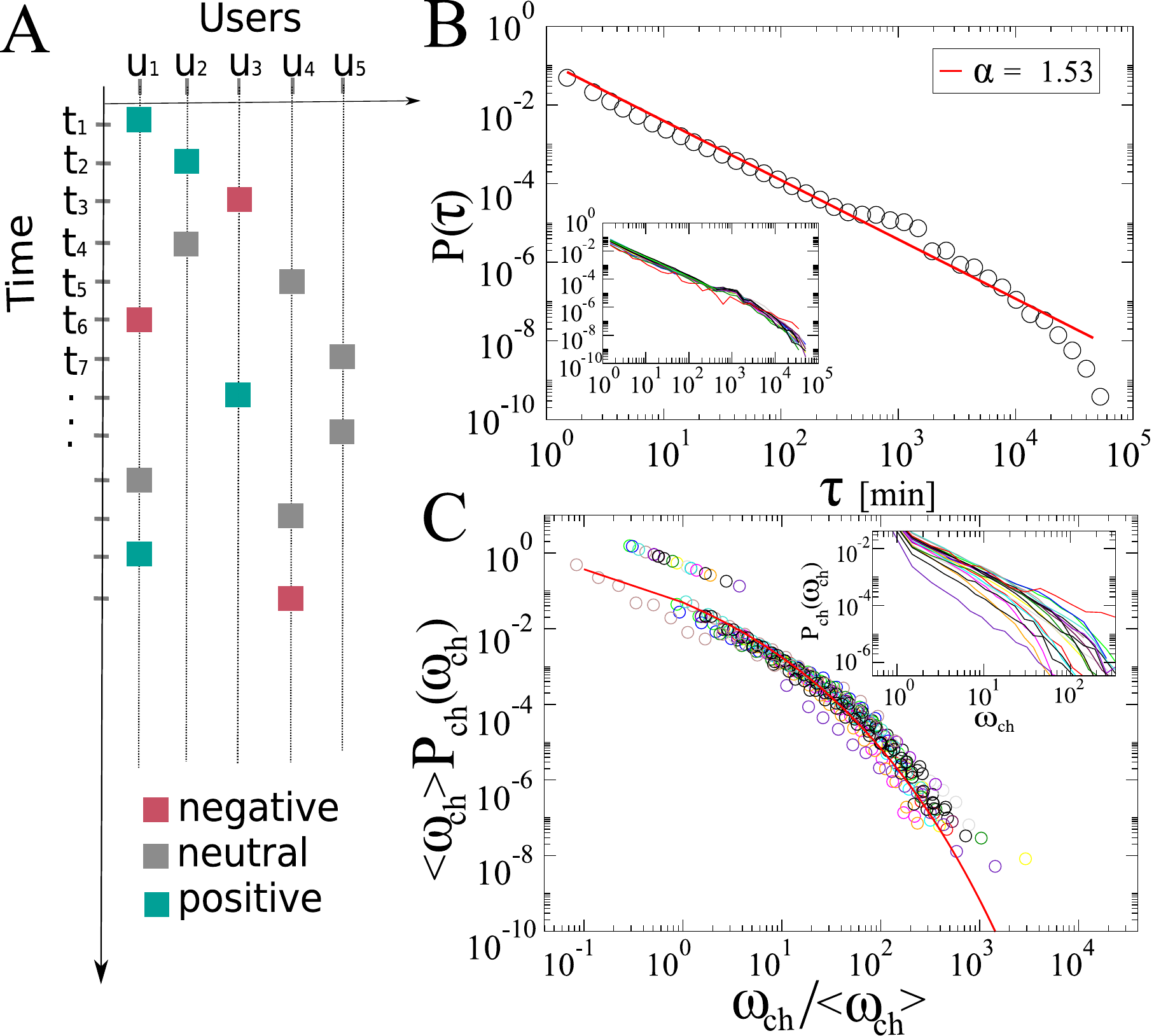}
 \caption{Communication activity over an IRC channel. A) Schematic
   evolution of a conversation in an IRC channel. At every time step, a
   user enters a post expressing a positive, negative, or neutral
   emotion. B) Probability distribution of the user activity over all the
   IRC channels. The activity is expressed as the time interval $\tau$
   between two consecutive posts of the same user. Inset: Probability
   distribution of the user activity for individual IRC channels. The
   time is measured in minutes. C) Scaled probability distribution of the
   time interval $\omega_{\rm ch}$ between consecutive posts entered in
   all the 20 IRC channels. The solid line represents stretched
   exponential fit to the data. Inset: Probability distribution of the
   time interval $\omega_{\rm ch}$ between consecutive posts entered in
   all the 20 IRC channels without rescaling. The time is measured in
   minutes.}
 \label{fig:Temporal}
\end{figure}

An IRC channel is always active, and enables the real time exchange of
posts among users about a specific topic.  User interaction is
instantaneous, the post written by user $u_{1}$ is immediately visible to
all other users logged into this channel, and user $u_{2}$ may reply
right away.  Fig.~\ref{fig:Temporal} illustrates the dynamics in such a
channel.  As time evolves new users may enter, others may leave or stay
quiet until they write follow-up posts at a later time.

To characterize these activity patterns, we analyzed the waiting-time, or
inter-activity time distribution $P(\tau)$, where $\tau$ refers to the
time interval between two consecutive posts of the \emph{same} user in
the same channel and ask about the average response time.  We find that
$\tau$ is power-law distributed $P(\tau)\sim \tau^{-\alpha}$ with some
cut-off (Fig.~\ref{fig:Temporal}B), with an exponent $\alpha=1.53\pm
0.02$. The fit is based on the maximum likelihood approach proposed in
\cite{Clauset2009} and the power-law nature of the distribution could not
be rejected ($p=0.375$).

This finding (a) is inline the power-law distribution already found for
diverse human activities
~\cite{Oliveira2005,Barabasi2005,Malmgren2008,Rybski2009,Grinstein2008,Crane2010}
and (b) classifies the communication process as belonging to the regime
where posts arrive faster than they can be processed. We note that for
$\alpha <2$, no average response time is defined (which would have been
the case, however, for the highly attentive regime). Further, we observe
in the plot of Fig.~\ref{fig:Temporal}B a slight deviation from the
power-law at a time interval of about one day, which shows that some
users have an additional regularity in their behavior with respect to the
time of the day they enter the online discussion. Such deviations were
usually treated as power-laws with an exponential cut-off, and can even
be explained based on simple entropic arguments
\cite{Baek2011,Adamic2011}. However, because of the ``bump" around the
one day time interval, our distribution also seems to provide further
evidence to the bi-modality proposed in \cite{Wu2010}. We should note,
however, that the tail is better fitted by a log-normal distribution
(KS=0.136) rather than an exponential (KS=0.190) or a Weibull (KS=0.188)
one (again using the maximum likelihood methodology described in
\cite{Clauset2009}) as shown in Fig.~\ref{fig:Temporal}B. Here, KS stands
for the Kolmogorov-Smirnov statistical test; the smaller this number, the
better the fit.

We now focus on an important difference between online chats and
previously studied forms of communication, such as mail or email
exchange, which mostly involve two participants.  Due to the collective
nature of chats, a chatroom automatically aggregates the posts of a much
larger amount of users, which allows us to study their collective
temporal behavior. If $\omega$ denotes the time interval between two
consecutive posts in the same channel independent of any user (also
denoted as inter-event time, and to be distinguished from the
inter-activity time characterizing a single user), we find that the
distribution $P(\omega)$ is is still fat-tailed, but does not follow a
power-law. Interestingly, the time interval between posts significantly
depends on the topic discussed in the channel (Inset of
Fig.~\ref{fig:Temporal}C). Some "hot" topics receive posts at a shorter
rate than others, which can be traced back to the different number of
users involved into these discussions. Specifically, we find that the
average inter-event time $\mean\omega_{\rm ch}$ depends on the amount of
users in the conversation and becomes smaller for more popular channels,
as one would expect.

If we rescale the channel dependent inter-event distribution
$P_{\mathrm{ch}}(\omega)$ using the average inter-event time
$\mean\omega_{\rm ch}$ per channel and plot $\mean{\omega_{\rm ch}}P_{\rm
  ch}(\omega_{\rm ch})$ versus $\omega_{\rm ch}/\mean{\omega_{\rm ch}}$,
we find that all the curves collapse into one master curve
(Fig.~\ref{fig:Temporal}C).  The general scaling form that we used is
$P(\omega)=(1/<\omega>)F(\omega/<\omega>)$, where F(x) is independent of
the average activity level of the component, and represents a universal
characteristic of the particular system. Such scaling behavior was
reported previously in the literature describing universal patterns in
human activity \cite{Candia2008}. We fit this master curve by a stretched
exponential~\cite{Altmann2005,Bunde2005,Wang2006}
\begin{equation}
  P(\omega)=\frac{a_{\gamma}}{\mean{\omega}} 
  e^{-\beta_{\gamma}\left(\frac{\omega}{\mean{\omega}}\right)^{\gamma}} 
\end{equation}
where the stretched exponent $\gamma$ is the only fit parameter, while
the other two factors $a_{\gamma}$ and $\beta_{\gamma}$ are dependent on
$\gamma$~\cite{Altmann2005}.  A histogram of the $\gamma$ values across
the 20 channels is shown in Supplementary Figure S2. Using only the
regression results with $p<0.001$ we find that the mean value of the
stretched exponents is $\mean{\gamma}=0.21\pm0.05$.

We note that stretched exponentials have been reported to describe the
inter-event time distribution in systems as diverse as
earthquakes~\cite{Bunde2005} and stock markets~\cite{Wang2006}. These
systems commonly exhibit long range correlations which seem to be the
origin of the stretched exponential inter-event time distributions
\cite{Altmann2005}. Long range correlations have also been reported in
human interaction activity \cite{Rybski2009, Rybski2011}, and we tested
their presence in the temporal activity over IRC communication. As shown
in the Supplementary Figure S3, we verified the existence of long range
correlations in the conversation activity. We found that the decay of the
autocorrelation function of the inter-event time interval between
consecutive posts within a channel is described by a power-law
\begin{equation}
  C(\Delta t) \sim (\Delta t)^{-\nu_{\omega}}
\end{equation}
with exponent $\nu_{\omega}\simeq0.82$. In addition, we applied the
Detrended Fluctuation Analysis (DFA) technique \cite{Peng1994}, described
in detail in the Methods section, and we found a Hurst exponent value,
$H_{\omega}\simeq 0.6$, which is well in agreement with the scaling
relation $\nu_{\omega} = 2-2H_{\omega}$. For a more detailed discussion
about scaling relations, and memory in time series please refer to
\cite{Kantelhardt2009}.

In conclusion, our analysis of user activities have revealed a universal
dynamics in online chatting communities which is moreover similar to
other human activities. This regards (a) the temporal activity of
individual users (characterized by a power-law distribution with exponent
$3/2$) and (b) the inter-event dynamics across different channels, if
rescaled by the average inter-event time (characterized by a stretched
exponential distribution with just one fit parameter). We will use these
findings as a point of departure for a more in-depth analysis -- because
obviously the essence of online communication in chatrooms, as compared
to other human activities, is not really covered. From the perspective of
activity patters, there is not so much new here, which leads us to ask
for other dimensions of human communication that could reveal a
difference.

\subsection*{Emotional expression patterns}

Human communication, in addition to the mere transmission of information,
also serves purposes such as the reinforcement of social bonds. This
could be one of the reasons why human languages are found to be biased
towards using words with positive emotional charge \cite{Garcia2011b}.
Humans, from the early stages of our lives, develop an affective
communication system that enables us to express and regulate
emotions\cite{Tronick1989}. But emotions are also the mediators of our
consumer responses to advertising~\cite{Holbrook1987}, and many
scientists acknowledge their importance in motivating our cognition and
action~\cite{Izard2010}.  However, despite the increasing time we spend
online, the way we express our emotions in online communities and its
impact on possibly large amounts of people is still to be explored.

\begin{figure}[h]
 \centering
% Figure 2
 \includegraphics[scale=0.6]{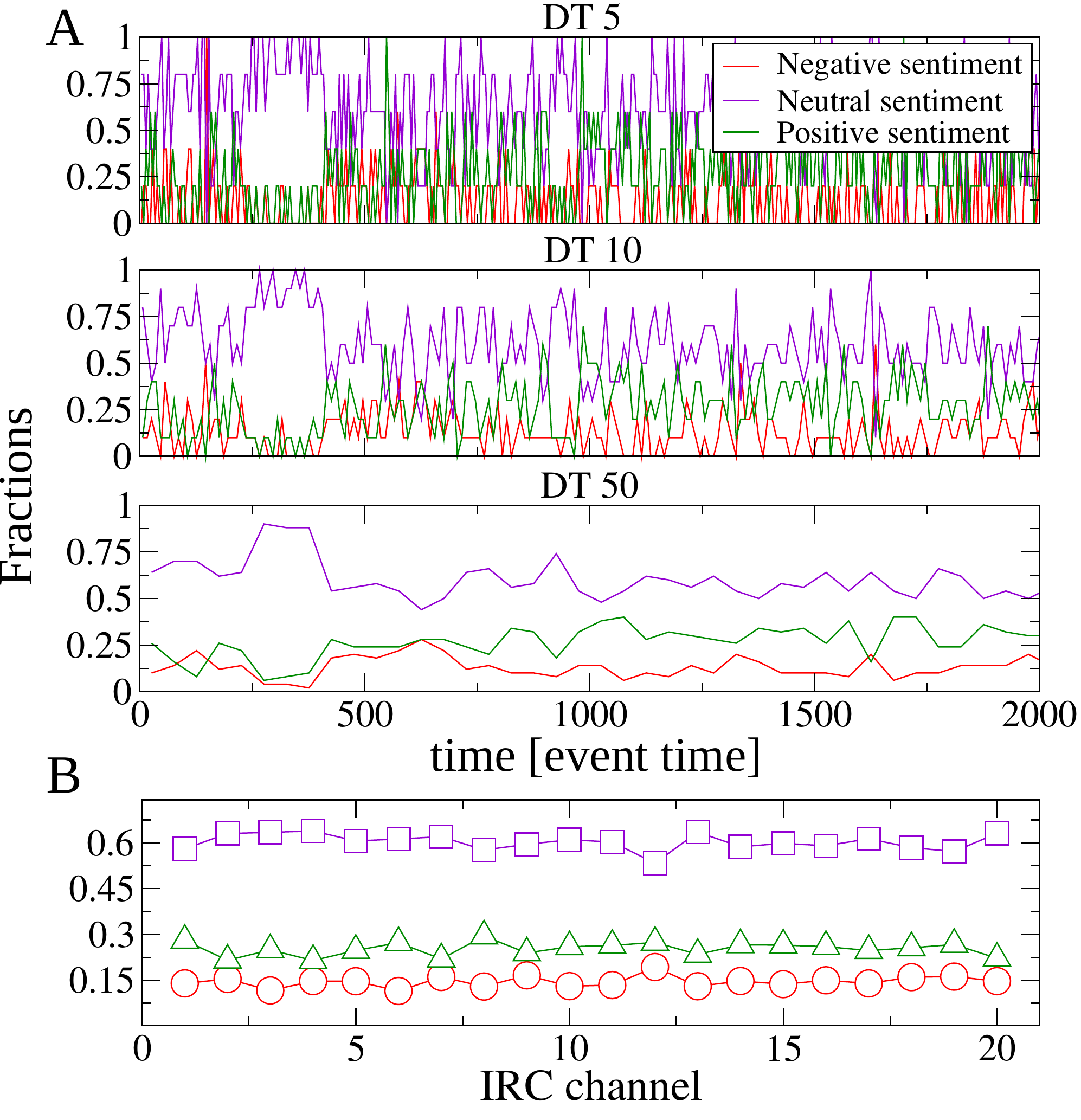}
 \caption{Emotional expressions over different time scales. A) Fraction
   of expressions with negative, neutral, and positive emotion values
   under different time scales for one channel. B) Fraction of
   expressions with negative, neutral, and positive emotion values for
   the 20 IRC channels.}
 \label{fig:Frac}
\end{figure}

\begin{figure}[h]
 \centering
% Figure 3
 \includegraphics[scale=0.46]{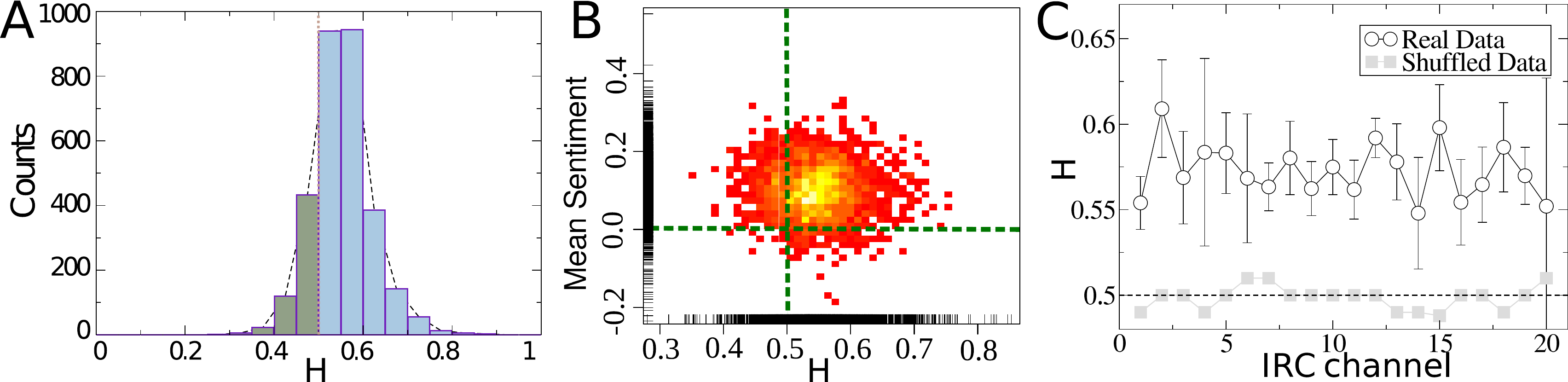}
 \caption{Hurst exponents and emotional persistence. A) Hurst exponents
   ($H$) of the emotion expression of individual users, obtained using
   the DFA method. Only users contributed more than 100 posts were
   considered, and we used the exponents obtained with fitting quality
   $R^2>0.98$. B) Hurst exponent ($H$) versus the mean emotion polarity
   expressed by individual users, again only from users who contributed
   more than 100 posts. C) Hurst exponents ($H$) of the emotions
   expressed in the 20 IRC channels. The values are averages of the Hurst
   exponents obtained from 10 different segments of the same channel, and
   the error bars show the standard deviation. The horizontal dashed line
   shows the expected value for random time series ($H=0.5$), and the
   gray squares show the value obtained from shuffling the real time
   series to destroy any correlations. The difference in exponents of the
   real and the shuffled time series is statistically significant with
   $p<0.001$.}
 \label{fig:Hurst}
\end{figure}

Consequently, we are interested in the role of expressed emotions in
online chatting communities. Users, by posting text in chatrooms, also
reveal their emotions, which in return can influence the emotional
response of other users, as illustrated in Fig.~\ref{fig:Temporal}A. To
understand this emotional interaction, we carry out a sentiment analysis
of each post which is described in detail in the Methods section. This
automatic classification returns the valence $v$ for each post, i.e. a
discrete value $\{-1,0,+1\}$ that characterizes the emotional charge as
either negative, neutral, or positive.

Instead of using the real time stamp of each post as in the analysis of
the user activity, we now use an artificial time scale in which at each
(discrete) time step one post enters the discussion, so the number of
time steps equals the total number of posts.  We then monitor how the
total emotion expressed in a given channel evolves over time. We use a
moving average approach that calculates the mean emotional polarity over
different time windows. In Fig.~\ref{fig:Frac}A we plot the fraction of
neutral, negative and positive posts as a function of time, for different
sizes of the time window. While it is obvious that the emotional content
largely fluctuates when using a very small time window, we find that for
decreasing time resolution (i.e.  increasing time window) the fractions
of emotional posts settle down to an almost constant value around which
they fluctuate. From this, we can make two interesting observations: (i)
the emotional content in the online chats does not really change in the
long run (one should notice that times of the order $10^{3}$ are still
large compared to the time window $DT=50$ used), i.e. we observe
fluctuations that depend on the time resolution, but no "evolution"
towards more positive or negative sentiments. (ii) For the low
resolution, the fraction of neutral posts dominates the positive and
negative posts at all times. In fact there is a clear ranking where the
fraction of negative posts is always the smallest.  Both observations
become even more pronounced when averaging over the 20 IRC channels, as
Fig.~\ref{fig:Frac}B shows.

Our findings differ from previous observations of emotional communication
in blog posts and forum comments which identified a clear tendency toward
negative contributions over time, in particular for periods of intensive
user activity~\cite{Mitrovic2009,Chmiel2010}. Such findings suggest that
an increased number of negative emotional posts could boost the activity,
and extend the lifetime of a forum discussion. However, blog
communication in general evolves slower than e.g. online chats. Hence, we
need to better understand the role of emotions in real time Internet
communication, which obviously differs from the persistent and delayed
interaction in blogs and fora.

To further approach this goal, we analyse to what extend the rather
constant fraction of emotional posts in IRC channels is due to a
persistence in the emotional expressions of users. For this, we apply the
DFA technique ~\cite{Peng1994}, to the time series of positive, negative
and neutral posts. Since our focus is now on the user, we reconstruct for
every user a time series that consists of all posts communicated in any
channel, where the time stamp is given by the consecutive number at which
the post enters the user's record.  In order to have reliable statistics,
for the further analysis only those users with more than 100 posts are
considered (which are nearly 3000 users).  As the examples in the
Supplementary Figure S4 show, some users are very persistent in their
(positive) emotional expressions (even that they occasionally switch to
neutral or negative posts), whereas others are really antipersistent in
the sense that their expressed emotionality rapidly changes through all
three states.  The persistence of these users can be characterized by a
scalar value, the Hurst exponent $H$, (see the Material and Methods
Section for details) which is 0.5 if users switch randomly between the
emotional states, larger than 0.5. if users are rather persistent in
their emotional expressions, or smaller than 0.5 if users have strong
tendency to switch between opposite states, as the antipersistent time
series of Fig. S4 shows.

If we analyse the distribution of the Hurst exponents of all users, shown
in the histogram of Fig.~\ref{fig:Hurst}A, we find (a) that the emotional
expression of users is far from being random, and (b) that it is clearly
skewed towards $H>0.5$, which means that the majority of users is quite
persistent regarding their positive, negative or neutral emotions. This
persistence can be also seen as a kind of memory (or inertia) in changing
the emotional expression, i.e.  the following post from the same user is
more likely to have the same emotional value.

The question whether persistent users express more positive or negative
emotions is answered in Fig.~\ref{fig:Hurst}B, where we show a scatter
plot of $H$ versus the mean value of the emotions expressed by each user.
Again, we verify that the majority of users has $H>0.5$, but we also see
that the mean value of emotions expressed by the persistent users is
largely positive. This corresponds to the general bias towards positive
emotional expression detected in written expression
\cite{Garcia2011b}. The lower left quadrant of the scatter plot is almost
empty, which means that users expressing on average negative emotions
tend to be persistent as well. A possible interpretation for this could
be the relation between negative personal experiences and rumination as
discussed in psychology \cite{Rime2009}.  Antipersistent users, on the
other hand, mostly switch between positive and neutral emotions.

Are the more active users also the emotionally persistent ones?  In
Supplementary Figure S6 we show a scatter plot of the Hurst exponent
dependent on the total activity of each user. Even though the mean value
of $H$ does not show any such dependence, we observe large heterogeneity
on the values of $H$ for users with low activity. Furthermore, in
Supplementary Figure S7 we show that the Hurst exponent of a very active
user varies only slightly if we divide his time series into various
segments and apply the DFA method to these segments. Thus we can conclude
that active users tend to be emotionally persistent and, as most
persistent users express positive emotions, they tend to provide some
kind of positive bias to the IRC, whereas users occasionally entering the
chat may just try to get rid of some negative emotions.

This leads us to the question how persistent the emotional bias of a
whole discussion is. While Fig.~\ref{fig:Hurst}A has shown the
persistence with respect to the different users, Fig.~\ref{fig:Hurst}C
plots the persistence for the different channels, which each feature a
very different topic.  This persistence holds even even if we analyse
only certain segments of the channel, as it is shown in Supplementary
Figure S8.  So, we conclude that the persistence of the discussion per se
(which is different from the persistence of the users which can leave or
enter a arbitrary times) reflects a certain narrative memory. Precisely,
for each chat, we observe the emergence of a certain (emotional) "tone"
in the narration which can be positive, negative or neutral, dependent
the emotional expressions of the (majority of) persistent users.  If we
reshuffle these time series such that the same total number of positive,
negative, and neutral posts is kept, but temporal correlations are
destroyed, then the persistence is lost as well as Fig.~\ref{fig:Hurst}C
shows. We note that we could not find evidence of correlations using the
autocorrelation function of the emotion time series, while the observed
persistence in the fluctuations of user emotional expression, as captured
by the Hurst exponent is very robust.  This indicates that the chat
community assumes an emotional memory locally encoded in the current
messages (from the user perspective), while the size of the conversation
is too large to detect it through averaging techniques.

\subsection*{An agent-based model for chatroom users}
\label{sec:model}
 
\begin{figure}[h]
 \centering
% Figure 4
 \includegraphics[scale=1.8]{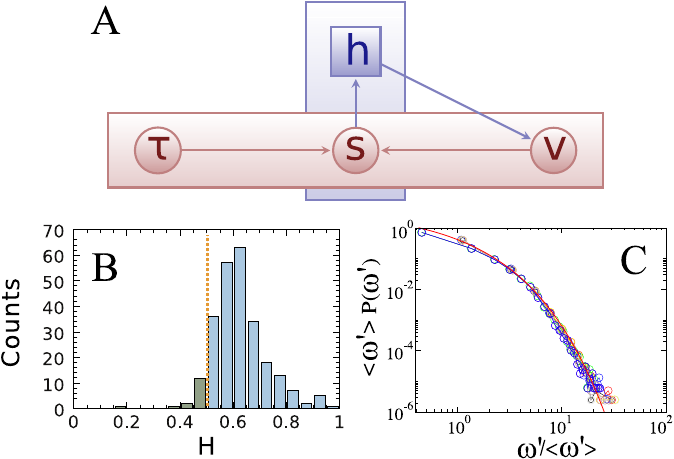}
 \caption{Modeling schema, and simulation results. A) Schematic feedback
   model: The horizontal layer represents the agent, the vertical layer
   the communication in the chatroom where posts are aggregated.  After a
   time lapse $\tau$, which follows the power-law distribution of
   Fig.\ref{fig:Temporal}B, the agents writes a post $s$ which
   implicitly expresses its emotions, $v$. Posts read in the chatroom
   feed back on the emotional state $v$ of the agent.  B) Hurst exponents
   for the individual behavior of agents in isolation with $A_v \in
   [0.2,0.5]$ and $\gamma_v \in [0.2,0.5]$. Only the exponents derived
   with fitting quality $R^2>0.9$ are considered. C) Scaled probability
   distribution of the time interval $\omega '$ between consecutive posts
   in 10 simulations of the model. Stretched exponential fit shows
   similar behavior to real IRC channel data.}
 \label{fig:Model}
\end{figure}

After identifying both the activity patterns, and the emotional
expression patterns of users in online chats, we setup an agent-based
model that is able to reproduce these stylized facts.  We start from a
general framework \cite{Schweitzer2010}, designed to model and explain
the emergence of collective emotions in online communities through the
evolution of psychological variables that can be measured in experimental
setups and psychological studies \cite{Kappas2011, Kuster2011}. This
framework provides a unified approach to create models that capture
collective properties of different online communities, and allows to
compare the different emotional microdynamics present in various types of
communication. The case of IRC channel communication is of particular
interest because of its fast and ephemeral nature. Thus, we have designed
a model for IRC chatrooms, as shown in Fig.~\ref{fig:Model}A. The agents
in our model are characterized by two variables, their emotionality, or
valence, $v$ which is either positive or negative and their activity, or
arousal, which is represented by the time interval $\tau$ between two
posts $s$ in the chatroom.  The valence of an agent $i$, represented by
the internal variable $v_i$, changes in time due to a superposition of
stochastic and deterministic influences \cite{Schweitzer2010,
  Schweitzer2003}: \begin{equation} \dot{v_i} = -\gamma_v v_i + b *
  (h_{+} - h_{-}) * v_{i} + A_{v} \xi_{i} \label{eq:valence} \end{equation}
The stochastic influences are modeled as a random factor $A_{v} \xi_i$
normally distributed with zero mean and amplitude $A_{v}$, and represent
all changes of the individual emotional state apart from chat
communication. The deterministic influences are composed of an internal
decay of parameter $\gamma_{v}$, and an external influence of the
conversation. The change in the valence caused by the emotionality of the
field $(h_{+} - h_{-})$ is measured in valence change per time unit
through the parameter $b$.  Previous models under the same framework
\cite{Schweitzer2010, Garcia2011} had an additional saturation term in
the equation of the valence dynamics.  This way the positive feedback
between $v$ and $h$ was limited when the field was very large. But, as we
show in Fig. \ref{fig:Frac}, chatrooms do not show the extreme cases of
emotional polarization observed in other communities. Thus, we simplify
the dynamics of the valence without using any saturation terms, since a
large imbalance between $h_{+}$ and $h_{-}$ is unrealistic given our
analysis of real IRC data.

In general, the level of activity associated with the emotion, known as
arousal, can be explicitly modeled by stochastic dynamics as well
\cite{Garcia2011}. Here, the activity of an agent is estimated by the
time-delay distribution that triggers the expression of the agent,
i.e. by the power-law distribution $P(\tau)\sim \tau^{-1.53}$ shown in
Fig. 1B.  Assuming that an agent becomes active and expresses its emotion
at time $t$, it will become active again after a period $\tau$.  The
agent then writes a post in the online chat the emotional content of
which is determined by its valence (see below). This information is
stored in an external field common for all agents, which is composed of
two components, $h_{-}$ and $h_{+}$, for negative and positive
information, and their difference measures the emotional charge of the
communication activity.  Since we are interested in emotional
communication, we assume that all neutral posts entered, or already
present, in a chatroom do not influence the emotions of the agents
participating to the conversation. Thus, the dynamics of the field is
influenced only by the amount of agents expressing a particular emotion
at a given time: $N_{+}(t) = \sum_i (1 - \Theta(-1*s_i))$ and $N_{-}(t) =
\sum_i (1 - \Theta(s_i))$, where $\Theta$ is the Heaviside step
function. Therefore, the time dynamics of the fields can be described as:
\begin{equation}
  \dot{h}_{\pm} = -\gamma_h h_{\pm} + c * N_{\pm}(t)
\end{equation} 
These two field components, $h_{+}$ and $h_{-}$, decay exponentially with
a constant factor $\gamma_h$, i.e. their importance decays very fast as
they move further down the screen (posts never disappear, but become less
influential).  Each field increases by a fixed amount $c$ from every post
stored in it. The values of the valence of the agents are changed by the
field components, as described by Eq.~\ref{eq:valence}.  In contrast with
traditional means of communication, online social media can aggregate
much larger volumes of user-generated information. This is why $h$ is
defined without explicit bounds. Chatrooms pose a special case to this
kind of communication, as they can contain large amount of posts but
limited amount of users. Most IRC channels have technical limitations for
the amount of users that can be connected at once, which in turn is
reflected in the total amount of posts present in the general
discussion. In our model, $h$ might take any value, but the empirical
activity pattern combined with the fixed size of the community
dynamically constraints it to limited values.

Whenever an agent creates a new post in an ongoing conversation, the
variable, $s_i$, obtain its value in the following way:
\begin{eqnarray}
   s_i = \left\{
     \begin{array}{rr}       
       -1 & \text{\quad if \quad} v_i < V_{-}\\
       +1 & \text{\quad if \quad} v_i > V_{+}\\
       0 & \text{\quad otherwise.}
     \end{array}
   \right.
\end{eqnarray} 
The thresholds $V_{-}$ and $V_{+}$ represent a limit value of the valence
that determines the emotional content of each post, and in general can be
asymmetric, as humans tend to have different thresholds for the
triggering of positive and negative emotional expression. Each action
contributes to the amount of information stored in the information field
of the conversation, increasing $h_{-}$ if $s=-1$ or $h_{+}$ if $s=+1$.

We emphasize that the way we model the agent behavior is very much in
line with psychological research, where emotional states are represented
by valence and arousal, following the dimensional representation of core
affect~\cite{Russell1980}.  The valence, $v$, represents the level of
pleasure experienced by the emotional state, while the arousal represents
the degree of activity induced by the emotional state, and determines the
moment when posts are created. Continuously the agent's valence relaxes to
a neutral state and is subject to stochastic influences, as show
empirically in \cite{Kuppens2010}. The effect of chatroom communication
on an agent's emotionality is modeled as an empathy-driven process
\cite{Preston2002} that influences the valence. In the valence dynamics
we propose in Eq. \ref{eq:valence}, agents perceive a positive influence
when their emotional state matches the one of the community, and a
negative one in the opposite case. When a post is created, its emotional
polarity is determined by the valence, as it was suggested by
experimental studies on social sharing of emotions \cite{Christophe1997,
  Rime2009}.

All the assumptions of our model are supported by psychological
theories. Parameter values and dynamical equations can be tested against
experiments in psychology, providing empirical validation for the
emotional microdynamics \cite{Kappas2011, Kuster2011}.  Furthermore, our
model provides a consistent view of the emotional behavior in chatrooms
leading to testable hypotheses that can drive future psychology research.

We performed extensive computer simulations using different parameter
sets (see supplementary material for details).  By exploring the
parameter space, we identified which parameter sets lead to similar
conversation patterns as observed in the real data.  We used such set to
simulate chats in 10 channels, and we analysed the agent's activity and
their emotional persistence. The results are shown in
Fig.~\ref{fig:Model}B, C. Specifically, we find that (a) the distribution
of Hurst exponents for individual agents is shifted towards positive
values similar to the one observed in real data, this way reproducing the
emotional persistence of the conversation without assuming any time
dependence between user expressions. Further, we reproduce (b) the
empirically observed stretched exponential distribution for the rescaled
time delays $\omega '$ between consecutive posts, without any further
assumptions.

We do note, however, that the stretched exponent, $\gamma=0.59\
(p<0.001)$, of the simulated distribution is different from real IRC
channels where it was $\gamma=0.21$, i.e. there is a faster decay in the
simulations. This could be explained by the fact that in the real chat
users usually write after they have read the previous post, i.e. there
are additional correlations in the times users enter a chat. These,
however, are not considered in the simulations, because agents post in
the chat at random after a given time interval $\tau$, i.e. there is no
additional coupling in posting times. Following the same approach as we
did for the real data, we calculated the Hurst exponent of the inter
simulated event time-series of the discussions. We found that
$H_{\omega'}=0.75$, however, we did not observe a power-law decay of the
autocorrelation function (see Supplementary Figure S12). This suggests
that the observed correlations are due to the power-law distributed
inter-event times used as input to our model, and it is inline with the
above discussion about the absence of coupling that also explains the
difference in the stretched exponents.

Eventually, we observe (c) the emotional persistence in the simulated
conversations. The mean Hurst exponent for the 10 simulated channels is
$H_{s} = 0.567\pm0.007$, whereas for the real IRC channels $H_{r} = 0.572
\pm 0.021$ was found. These results suggests that our agent-based model
reproduces qualitatively the emergence of emotional persistence in the
IRC conversation and thus, based on all findings, is able to capture the
essence of emotional influence between users in chatrooms.

\section*{Discussion}

We started with the question to what extent human communication patterns
change on the Internet. To answer this, we used a unique dataset of
online chatting communities with about 2.5 million posts on 20 different
topics. Our analysis considered two different dimensions of the
communication process: (a) activity, expressed by the time intervals
$\tau$ at which users contribute to the communication, and $\omega$ at
which consecutive posts appear in a chat, and (b) the emotional
expressions of users. With respect to activity patterns we did not find
considerable differences between online chatrooms and other previously
studied forms on online and offline communication. Specifically, both the
inter-activity distribution of users and the inter-event distribution of
posts followed the known distributions. Thus, we may conclude that humans
do not really change their activity patterns when they go
online. Instead, these patterns seem to be quite robust across online and
offline communication.

The picture differs, however, when looking at the emotional expressions
of users. While we cannot directly compare our findings on emotional
persistence to results about offline communication, we find differences
between online chatrooms and other forms online communication, such as
blogs, fora. While the latter could be heated up by negative emotional
patterns, we observe that online chats, which are instantaneous in time,
very much follow a balanced emotional pattern across all topics (shown in
the emotional persistence of the channels), but also with respect to
individual users, which are in their majority quite persistent in their
emotional expressions (mostly positive ones).

This observation is indeed surprising as online chats are mostly
anonymous, i.e. users do not reveal their personal identity. However,
they still seem to behave according to certain social norms, i.e. there
is a clear tendency to express an opinion in a neutral to positive
emotional way, avoiding direct confrontations or emotional debates.  One
of the reasons for such behavior comes from the "repeated interaction"
underlying online chats. As the daily "bump" the activity patterns also
suggest, most users return to the online chats regularly, to meet other
users they may already know. This puts a kind of social pressure on their
behavior (even in an unconscious manner) to behave similar to offline
conversations. In conclusion, we find that the online communication
patters do not differ much from common offline behavior if a repeated
interaction could be assumed.

Eventually, we argue that the emotional persistence found is indeed
related to the nature of human conversations. After all, the correlations
shown in the emotional expressions of different users indicate that there
is some form of emotional sharing between participants. This suggests the
presence of social bonds among users in the chatroom \cite{Rime2009} and
confirms similarities between online and offline communication.

The fact that we could reveal patterns of emotional persistence both in
users and in topics discussed, does not mean that we also understand
their origin. One important step towards this "microscopic" understanding
is provided by our agent-based model of emotional interactions in
chatrooms. By using assumptions about the agent's behavior which are
rooted in research in psychology, we are able to reproduce the stylized
facts of the chatroom conversation, both for the activity in channels and
for the emotional persistence. Specifically, our model allows us to test
hypotheses about the emotional interaction of agents against their
outcome on the systemic level, i.e. for the chatroom simulation. This
helps to reveal what kind of rules are underlying the online behavior of
users which are hard to access otherwise.

\section*{Methods}

\subsection*{Data collection and classification}

The data used in this article is based on a large set of public channels
from EFNET Internet Relay Chats ({\it http://www.efnet.org}), to which
any user can connect and participate in the conversation.  Based on the
assessment of the initially downloaded set of recordings, 20 IRC channels
were selected aiming to provide a large number of consecutive daily logs
with transcripts of vivid discussions between the channel participants,
measured in number of posts. The finally used data set contained
consecutive recordings for 42 days spanning the period from 04-04-2006 to
15-05-2006.

The general topics of discussions from the selected channels include:
music, sports, casuals chats, business, politics and topics related to
computers, operating systems or specific computer programs.  The IRC data
set contains 2,688,760 posts.  The total number of participants to all
this channels is 25,166. However, because some people participate to more
than one channel, the total number of unique participants is 20,441. On
average, the data set provides 3055 posts per day. In the recorded period
15 users created more than 10000 posts. The distribution of the user
participation i.e. the number of posts entered by every user, is shown in
Supplementary Figure S1. The mean of the distribution is 97 posts per
user, and as we can see from Fig. S1, it is skewed with most of the users
contributing only a small number of posts.

The acquired data was anonymized by substituting real user ids to random
number references. The text of each post was cleaned by spam detection
and substitution of URL links to avoid them from influencing the emotion
classification. The emotional content was extracted by using the
SentiStrength classifier \cite{Thelwall2010}, which provides two scores
for positive and negative content. Each score ranges from 1 to 5, and
changes with the appearance of emotion bearing terms from a lexicon of
affective word usage, specifically designed for this purpose.  Each word
of the lexicon has a value on the scale of -5 to 5 which determines the
strength of the emotion attached to it. The classifier takes into account
syntactic rules like negation, amplification and reduction, and detects
repetition of letters and exclamation signs as amplifiers. When one of
this patterns is detected, SentiStrength applies transformation rules to
the contribution of the involved terms to the sentence scores. It has
been designed to analyze online data, and considers Internet language by
detecting emoticons and correcting spelling mistakes.

The perception of emotional expression varies largely across humans, and
traditional accuracy metrics are not useful when there is lack of an
objective space. Human ratings of emotional texts have certain degree of
disagreement that needs to be considered by sentiment analysis in order
to have a valid quantification of emotions.  SentiStrength scores are
consistent with the level of disagreement between humans about how they
perceive written emotional expressions \cite{Thelwall2012}. This
classifier combines an emotion quantization of proved validity with a
high accuracy, and is considered the state of the art in sentiment
detection \cite{Kucuktunc2012}. Due to the short length of the posts in
chatrooms, we calculate a polarity measure by comparing the two different
scores of SentiStrength.  The sign of the difference of the positive and
negative scores provides an approximation to detect positive, negative
and neutral posts. The accuracy of this polarity metric was tested
against texts tagged by humans and messages including emoticons from
MySpace \cite{Paltoglou2010} and Twitter \cite{Thelwall2011}, which are
of a similar length to the ones in our chatroom data.  The data are
freely available for research purposes, and are provided as Supplementary
Material. Detailed information about their structure is provided in the
``Data section" of the Supplementary Information text.

\subsection*{Detrended Fluctuation Analysis}
\label{methods:dfa}

The method of Detrended Fluctuation Analysis (DFA)~\cite{Peng1994} is a
useful tool in revealing long-term memory and correlations in time
series~\cite{Bunde2005,Wang2006,Rybski2009}. The method maps the system
into a one-dimensional random walk, and enable us to compare the
properties of the real time series with the time series produced by the
random case.

The DFA analysis of a time series $x(t)$ with length $T$, which can be
divided into $N$ segments is performed as follows: First we integrate the
time series, by calculating the profile
$Y(t)=\sum_{t'}^{t}[x(t')-<x(t)>]$. Next, we divide the integrated time
series into $N$ boxes of equal length $\Delta t$. Each box has a local
trend, which in a first level approximation, can be fitted by a linear
function using least squares. We denote with $y_{\Delta t}(t)$ the $y$
coordinate of the straight line segments that represent the local trend
in each box, and we subtract this local trend from the integrated time
series $Y(t)$. Next we use the function
\begin{equation}
F(\Delta t)=\sqrt{\frac{1}{N}\sum_{k=1}^{N}[Y(k)-y_{\Delta t}(k)]^{2}}
\end{equation}
to calculate the root-mean-square fluctuation of the integrated and
detrended time series, and we characterize the relationship between the
average fluctuation $F(\Delta t)$, and the box size $\Delta t$.

Typically, $F(\Delta t)$ will increase with box size as $F(\Delta t)\sim
(\Delta t)^{H}$, which indicates the presence of power-law (fractal)
scaling. Therefore, the fluctuations can be characterized only by the
scaling exponent $H$ that is analogous to the Hurst
exponent~\cite{Hurst1951}, and it is calculated from the slope of the
line relating ${\rm log}F(\Delta t)$ to ${\rm log}\Delta t$. If only
short-range correlations (or no correlations) exist in the time series,
then it has the statistical properties of a random walk. Therefore
$F(\Delta t)\sim (\Delta t)^{1/2}$. However, in the presence of
long-range power-law correlations (i.e. no characteristic length scale)
$H \neq 1/2$. A value $H<1/2$ signals the presence of long range
anti-correlations, while a value $H>1/2$ signals the presence of long
range correlations (persistence).

%\section{References}

\section*{Acknowledgments}
This research has received funding from the European Community's Seventh
Framework Programme FP7-ICT-2008-3 under grant agreement no 231323
(CYBEREMOTIONS).

\section*{Author contributions}
A.G., D.G., and F.S. designed the research, performed the research,
analysed the data, and wrote the manuscript. M.S. collected the data and
analysed the data.

\section*{Additional Information}

\subsection*{Competing Financial Interests}
The authors declare no competing financial interests.

\pagebreak
\newpage

\setcounter{figure}{0}
\makeatletter 
\renewcommand{\thefigure}{S\@arabic\c@figure} 

\section*{Sypplementary Information}

\subsection*{Supplementary figures}

\begin{figure}
 \centering
%Figure S1
% \includegraphics[scale=0.35]{S.I.FIGs/All_UserParticipation}
 \includegraphics[scale=0.35]{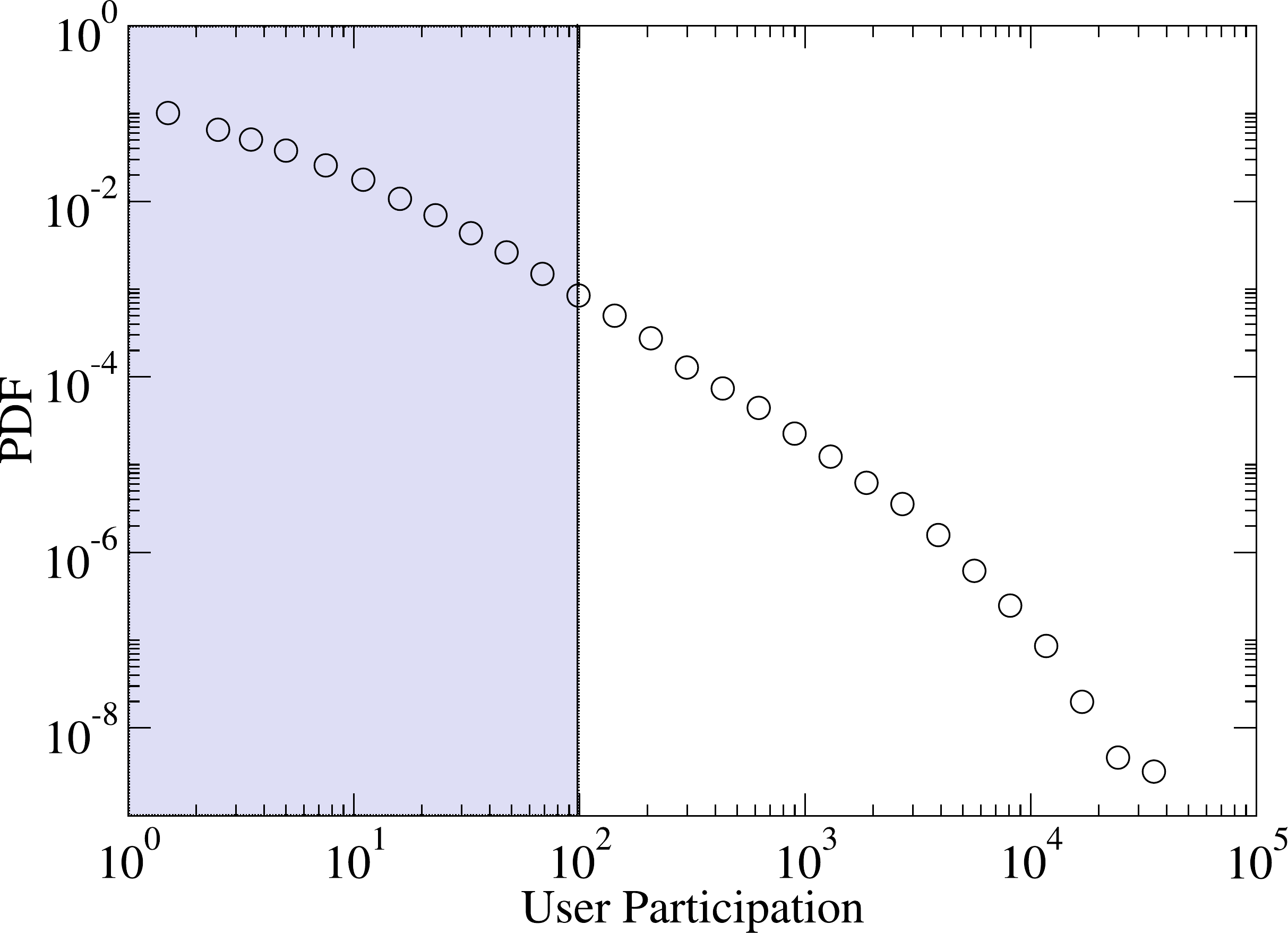}
 \caption{Distribution of the user participation in terms of the total
   number of posts entered by every user. The distribution is broad, and
   it is clear that most of the users contribute only a small number of
   posts. The shaded area shows the part of the user activity that is
   excluded from the DFA analysis in order to improve the statistical
   reliability of the results.}
 \label{Fig.S:UserParticipation}
\end{figure}

\begin{figure}
 \centering
%Figure S2
% \includegraphics[scale=0.65]{S.I.FIGs/StrechedExponents}
 \includegraphics[scale=0.65]{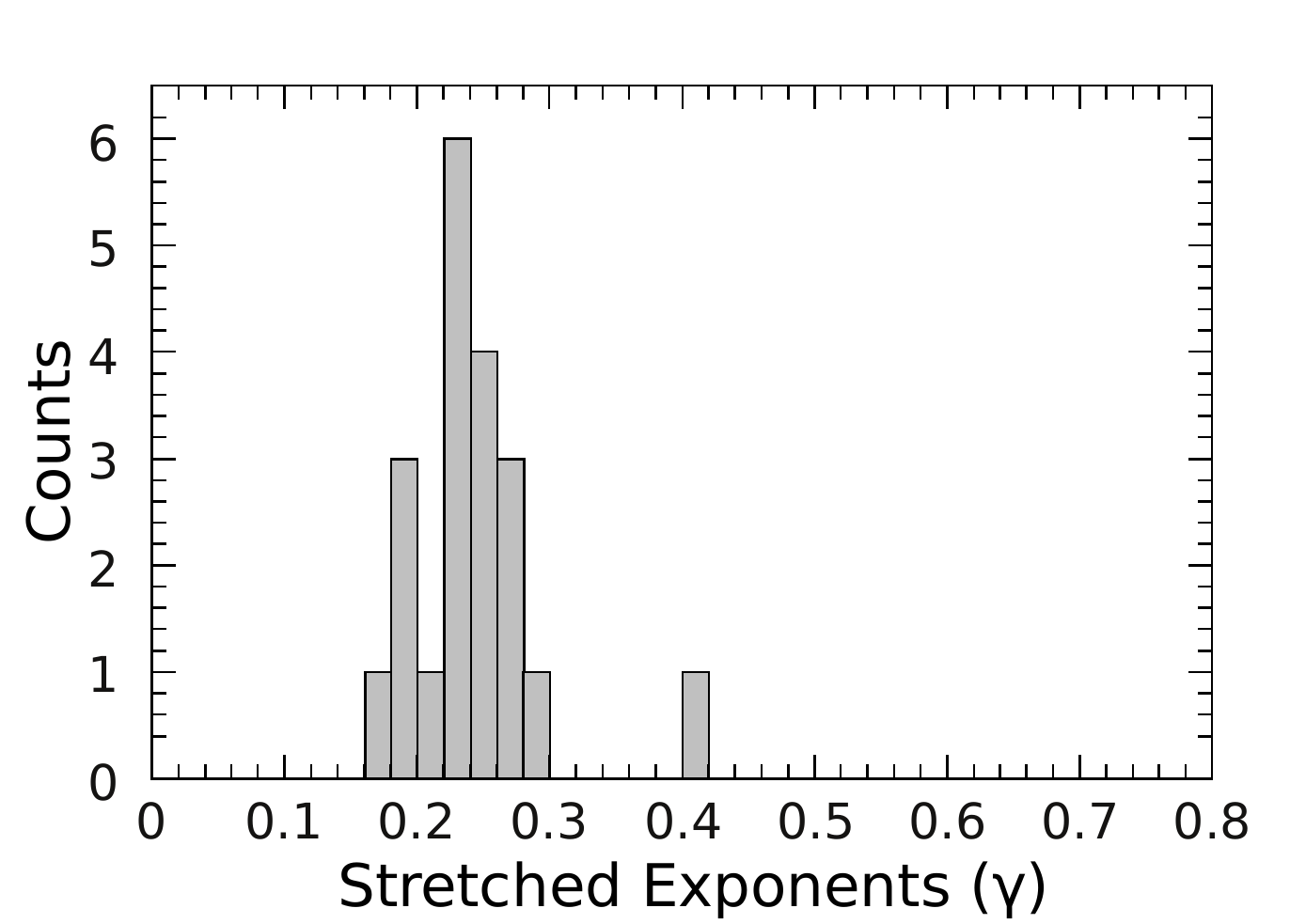}
 \caption{Histogram of stretched exponents obtained by fitting a
   stretched exponential function to the rescaled inter-event time of
   each individual channel separately. The exponents are concentrated
   around the mean value $\mean{\gamma}=0.21\pm0.05$, obtained using only
   the regression results with $p<0.001$, as explained in the text.}
 \label{Fig.S:S.Exp}
\end{figure}

\begin{figure}
 \centering
%Figure SXX1
%Figure S3
 \includegraphics[scale=0.35]{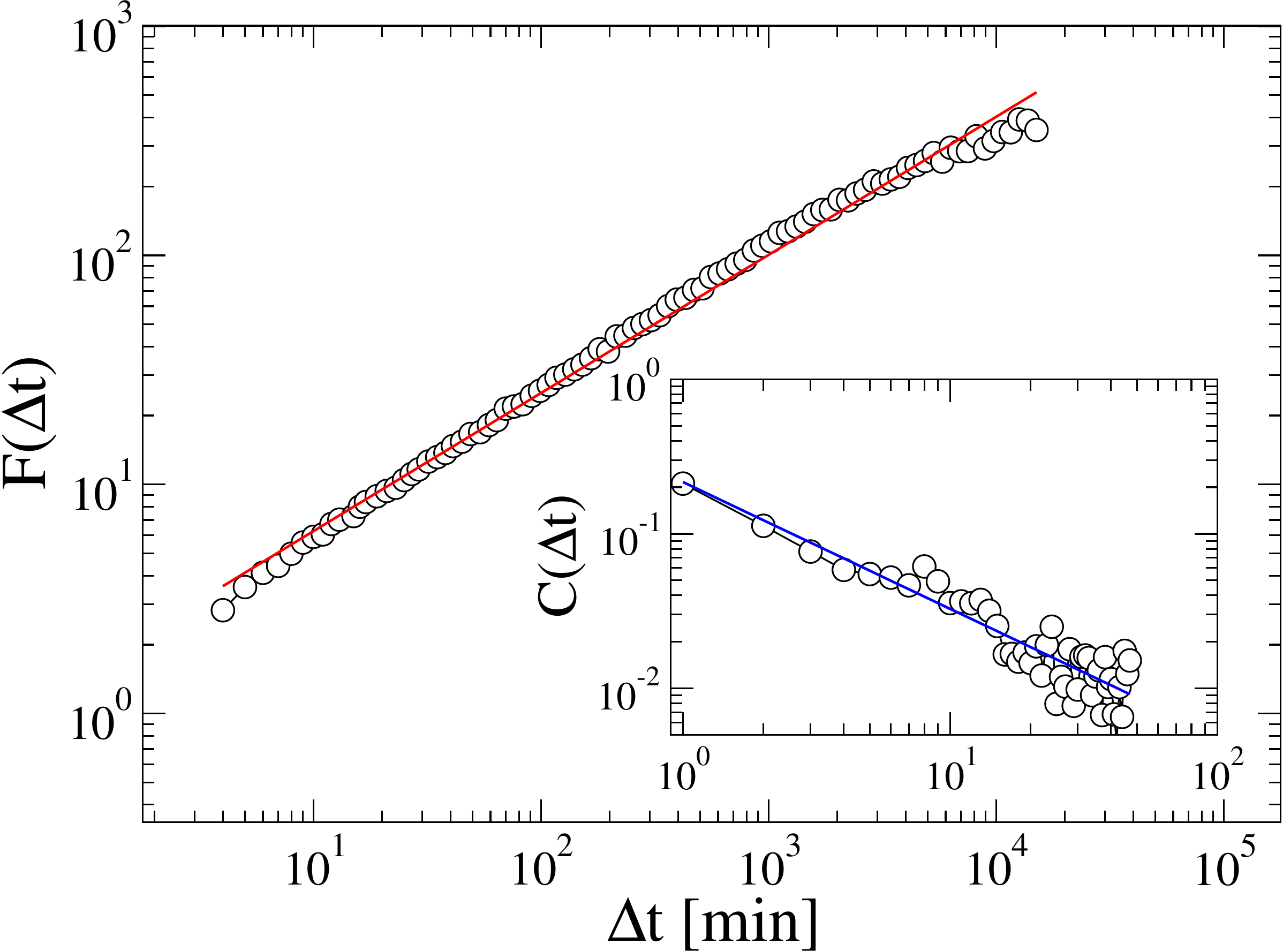}
 \caption{DFA fluctuation function calculated using the inter-event times
   of a real IRC channel. The Hurst exponent obtained is
   $H_{\omega}\simeq 0.6$, suggesting the existence of log term
   correlations in the time series. The origin of such correlations could
   be due to synchronized burst of activity leading to persistent
   dependencies over different time scales, or due to the broad
   distribution of inter-event times, or to a combination of both. The
   existence of dependencies in the activity is highlighted by a power
   law decaying autocorrelation function (Inset), with exponent
   $\nu_{\omega}\simeq 0.82$. The Hurst exponent is in scaling relation
   with the correlation exponent, given by $\nu_{\omega}=2-2H_{\omega}$
   \cite{Kantelhardt2009}.}
 \label{Fig.S:XX1}
\end{figure}

\begin{figure}
 \centering
%Figure S4
% \includegraphics[scale=0.35]{S.I.FIGs/Persistent-Antipersistent-TS}
 \includegraphics[scale=0.35]{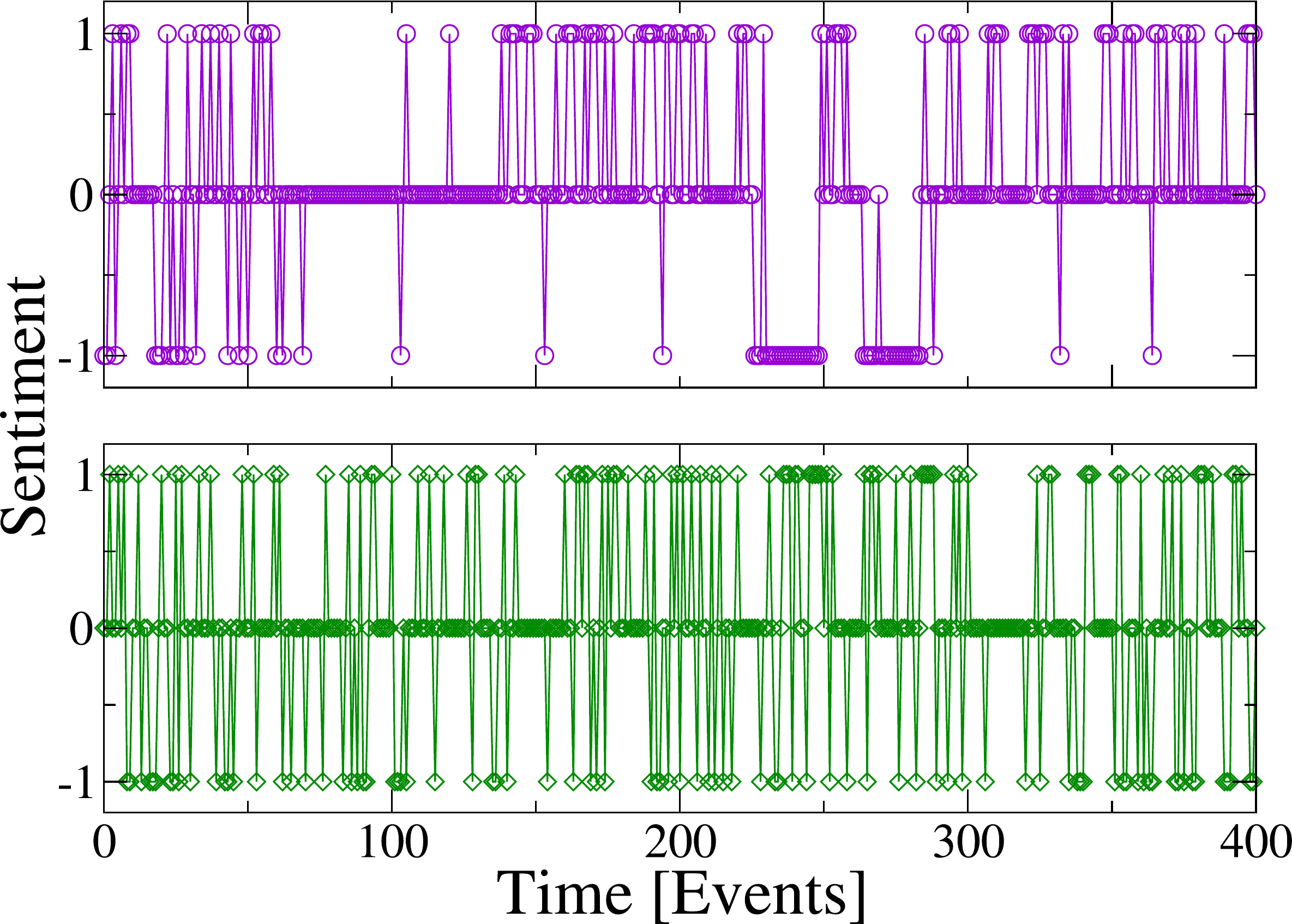}
 \caption{Time series showing examples of the sentiment expression for
   two real users. Top: An example of persistent sentiment time
   series. Bottom: An example of anti-persistent sentiment time series.}
 \label{Fig.S:P-A.TS}
\end{figure}

\begin{figure}
 \centering
%Figure S5
% \includegraphics[scale=0.35]{S.I.FIGs/Persistent-Antipersistent-DFA}
 \includegraphics[scale=0.35]{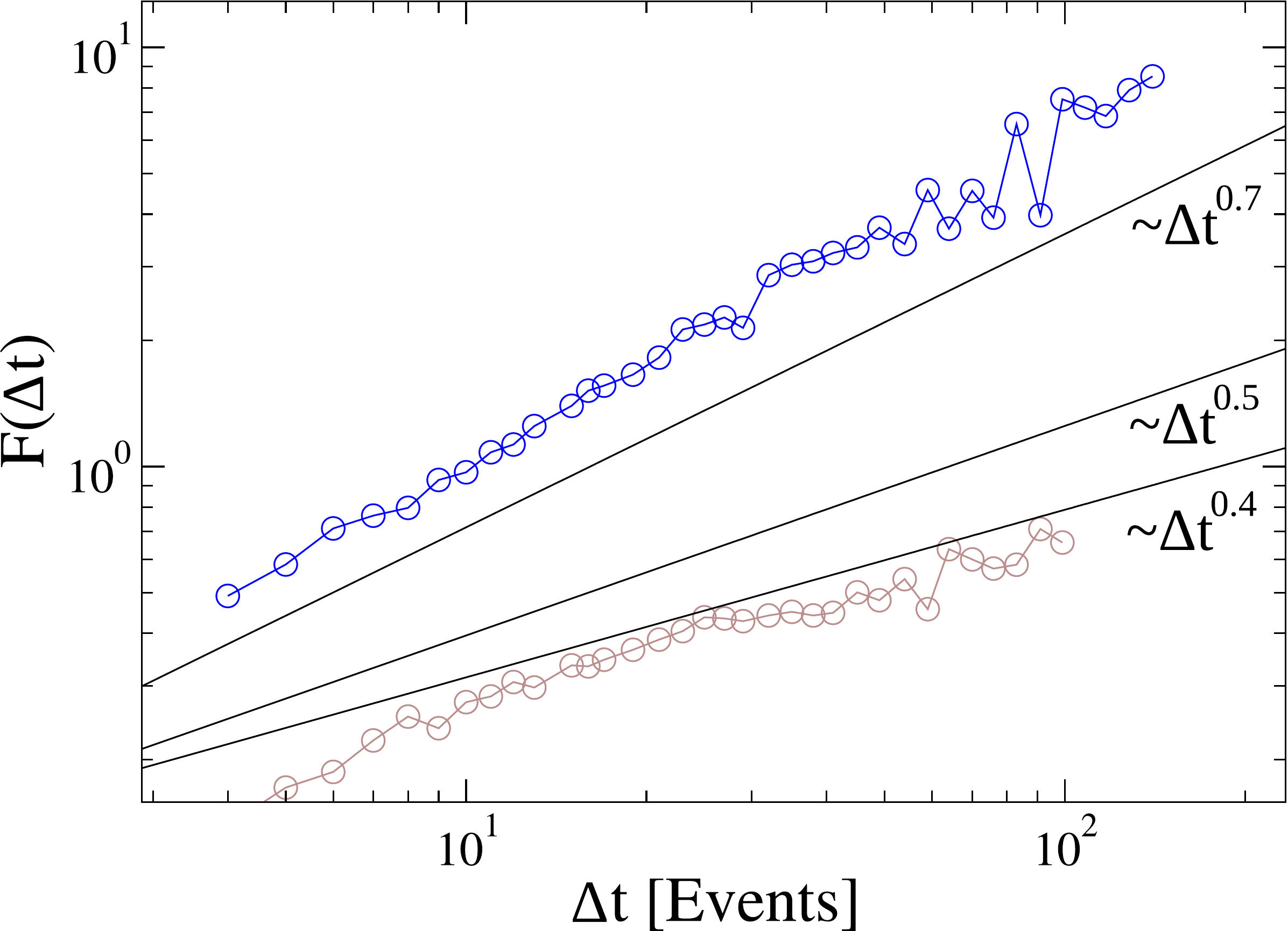}
 \caption{DFA fluctuation functions calculated for a persistent and an
   anti-persistent sentiment time-series. The solid lines are guides to
   the eye.}
 \label{Fig.S:P-A.TS.DFA}
\end{figure}

\begin{figure}
 \centering
%Figure S6
% \includegraphics[scale=0.65]{S.I.FIGs/HvsActivity}
 \includegraphics[scale=0.65]{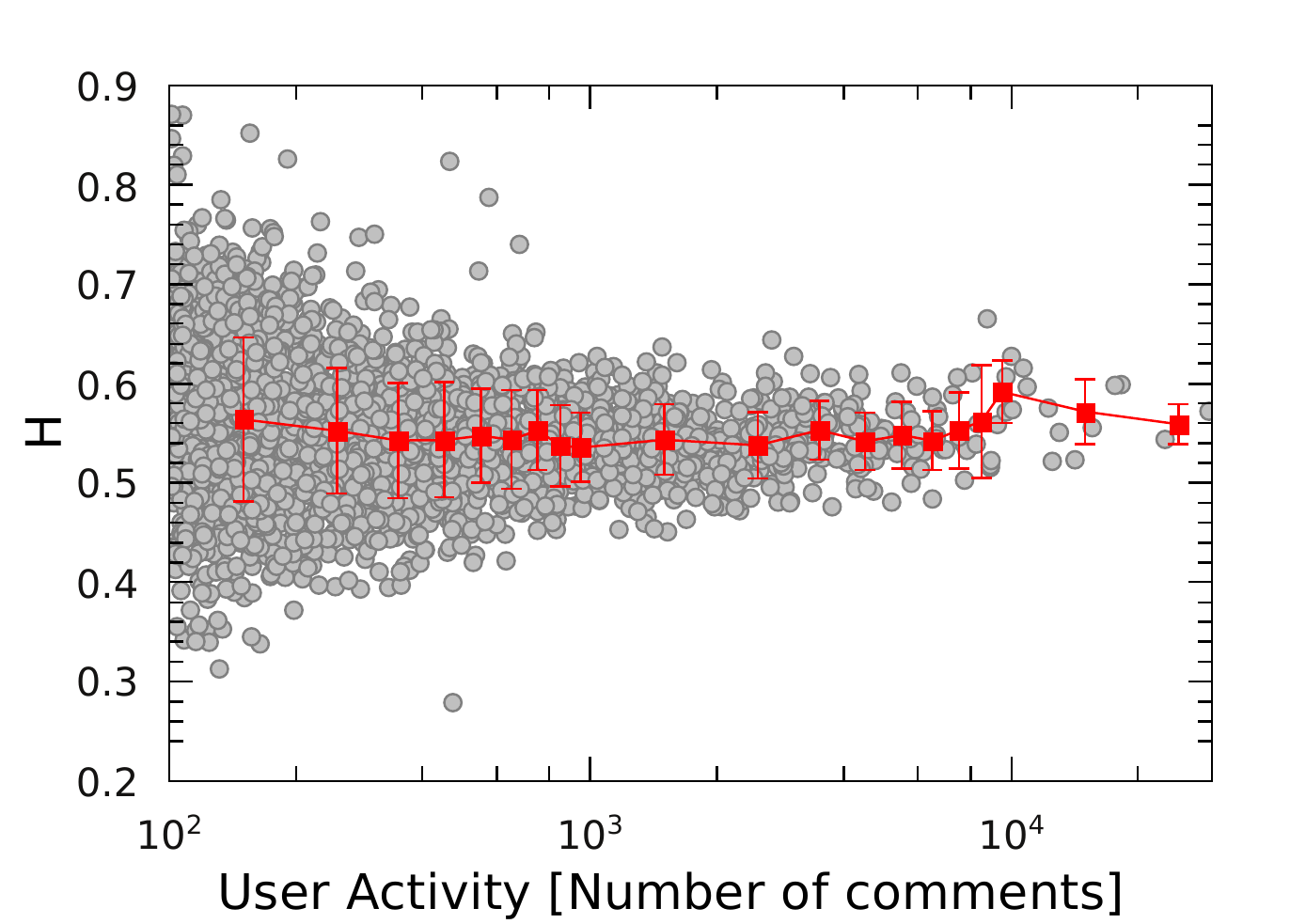}
 \caption{Dependence of the Hurst exponent on the total activity of each
   user. The mean value of H does not show any noticeable dependence on
   the activity but some large heterogeneity on the values of H for users
   with low activity is apparent.}
 \label{Fig.S:HvsActivity}
\end{figure}

\begin{figure}
 \centering
%Figure S7
% \includegraphics[scale=0.35]{S.I.FIGs/HvsSegments}
 \includegraphics[scale=0.35]{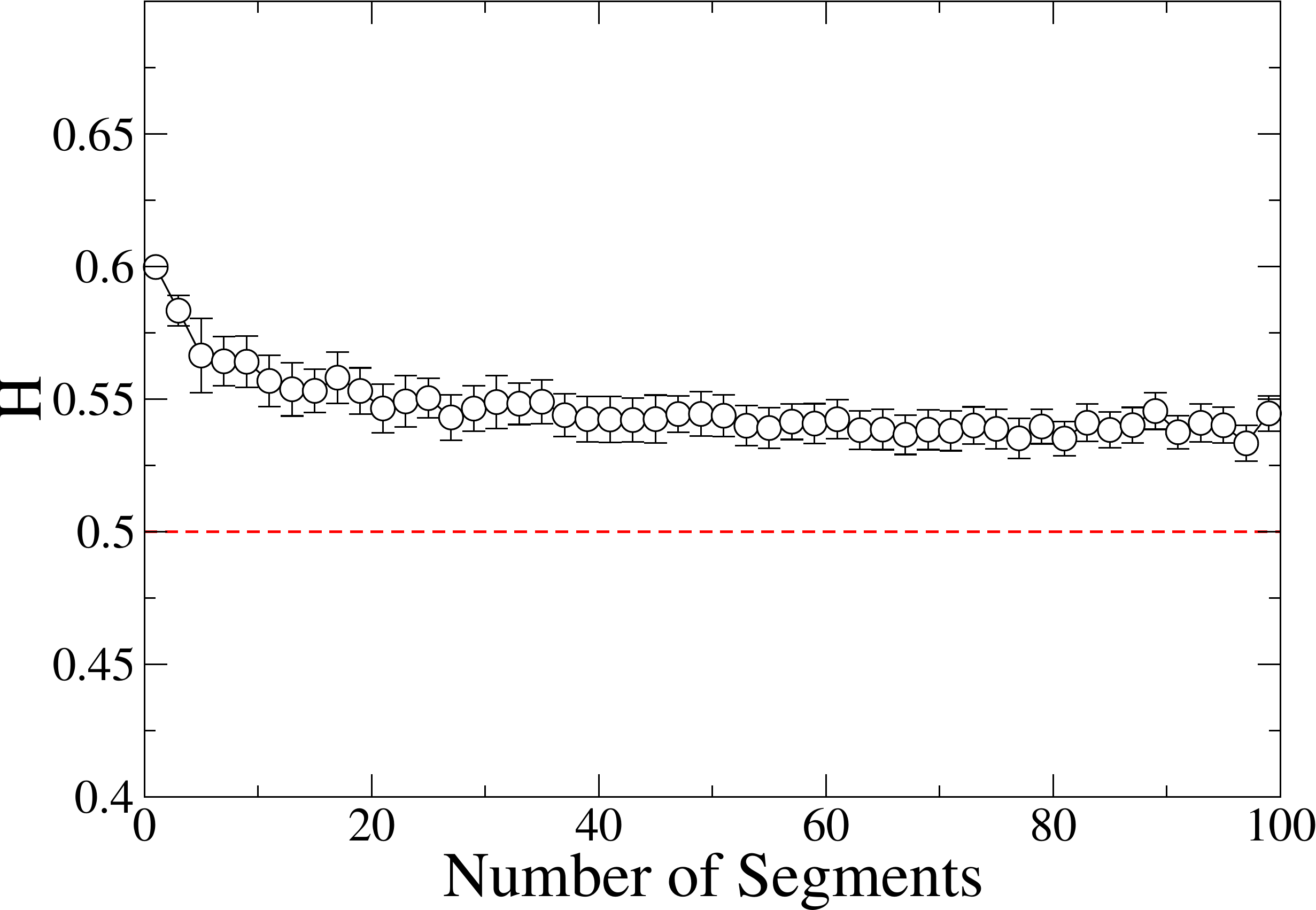}
 \caption{Dependence of the Hurst exponent on the length of the time
   series. We divide the expression time series of an active user into
   various segments and apply the DFA method to these segments. A small
   dependence on the length of the segments is observed, but the overall
   behavior of the user remains consistent. The error bars show the
   standard error of the mean. The total number of posts contributed by
   this user is 18.142, and the maximum number of segments we used was
   100 of length 181.}
 \label{Fig.S:HvsSegments}
\end{figure}

\begin{figure}
 \centering
%Figure S8
% \includegraphics[scale=0.35]{S.I.FIGs/Ch2-Segs-DFA}
 \includegraphics[scale=0.35]{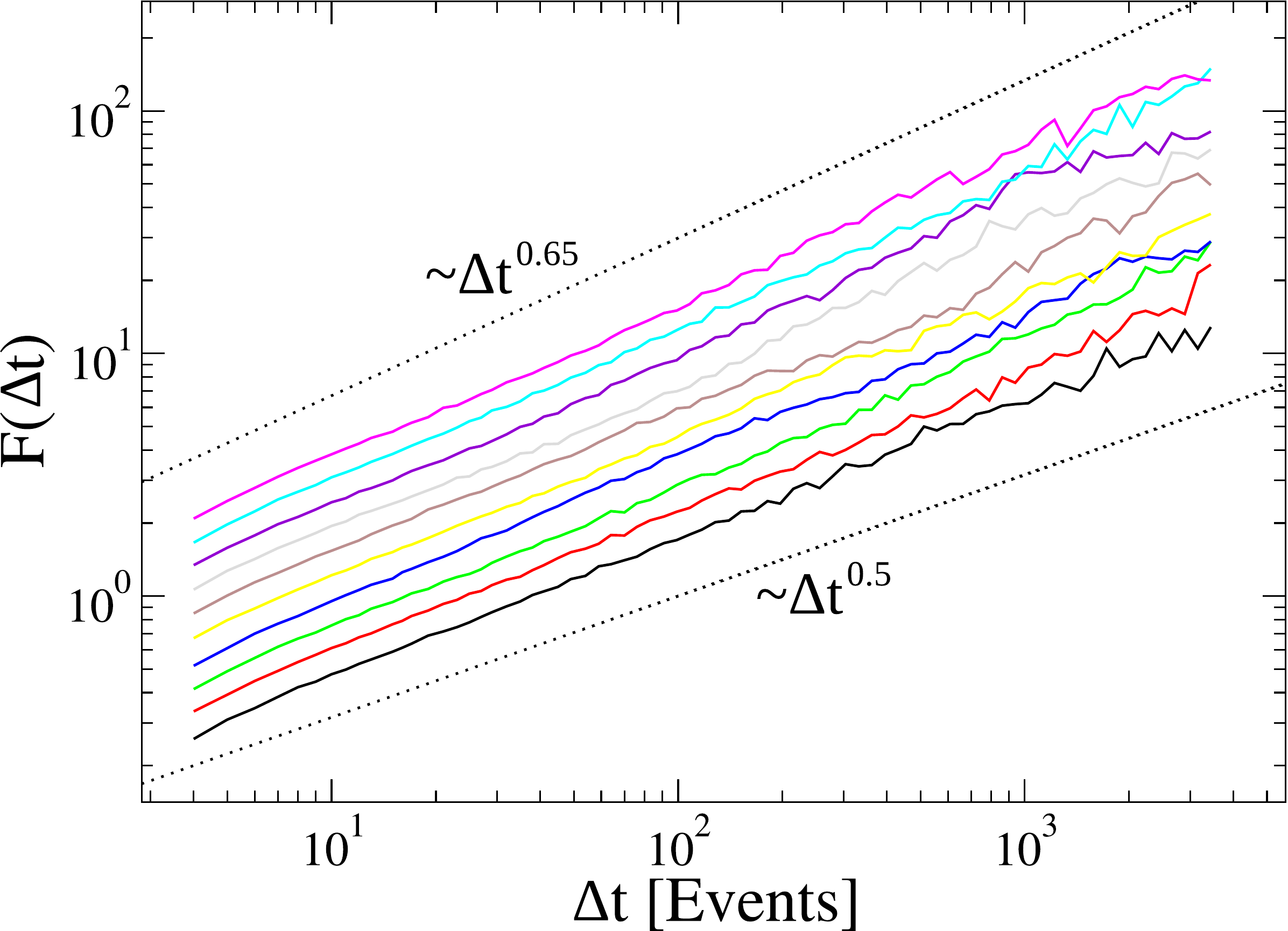}
 \caption{DFA fluctuation functions obtained for different segments of
   the time series describing the sentiment of a real IRC channel. It is
   clear that the persistence holds for all the segments analysed. The
   dashed lines are guides to the eye.}
 \label{Fig.S:Ch2SegsDFA}
\end{figure}

%Hereafter follows a set of supplementary figures :
\begin{itemize}
\item Supplementary Figure \ref{Fig.S:UserParticipation}: Distribution
  of the user participation. 
\item Supplementary Figure \ref{Fig.S:S.Exp}: Histogram of stretched 
  exponents.
\item Supplementary Figure \ref{Fig.S:XX1}: DFA and autocorrelation 
  analysis of real IRC channel activity
\item Supplementary Figure \ref{Fig.S:P-A.TS}: Example of persistent and
  anti-persistent 
  time series.
\item Supplementary Figure \ref{Fig.S:P-A.TS.DFA}: DFA fluctuation
  functions. 
\item Supplementary Figure \ref{Fig.S:HvsActivity}: Dependence of the
  Hurst exponent on the total activity 
   of each user.
\item Supplementary Figure \ref{Fig.S:HvsSegments}: Dependence of the
  Hurst exponent on the length of the 
   time series.
\item Supplementary Figure \ref{Fig.S:Ch2SegsDFA}: DFA fluctuation
  functions for different segments of 
   the time series.
\end{itemize}

\newpage

\newpage

\clearpage
\newpage

\subsection*{Data}

Our data set includes the annotated and anonymized logs from public
Internet Relay Chat (IRC) channels of EFNET
\footnote{http://www.efnet.org/} . In particular, the data consists of
consecutive daily recordings for 20 IRC channels for the period:
4-04-2006 - 17-05-2006. The general topics of discussions on these
channels, as indicated by the IRC channel names, include: music, sports,
casuals chats, business, politics and topics related to computers,
operating systems or specific computer programs. The data were anonymized
by substituting the real userIDs and the IRC channel names with generic
number references. Subsequently, the data were annotated according to:

\begin{itemize}
\item{\bf Sentiment classification}\\
  As described in the "Methods section" of the article, our emotional
  classification is based on the SentiStrength classifier
  \cite{Thelwall2010}, which provides two scores for positive
  (called positiveArousal) and negative (called negativeArousal)
  content. For example, the text {\it``I love you"} according to SentiStrenth has
  positiveArousal 3 and negativeArousal -1, while the text {\it ``I'm very
    sad"} has positiveArousal 1 and negativeArousal -5. \\ From these two
  scores, we calculate a polarity measure (called sentimentClass)
  using the sign of the difference of the positive and negative
  scores. This measure takes the values +1, -1, and 0, and it provides an
  approximation to detect positive, negative and neutral posts
  respectively. Under this approach, the sentimentClass of the first text
  would be +1 indicating a positive text, while the sentimentClass of the
  second text would be -1 indicating a negative text.

\item {\bf Affective, cognitive and linguistic categories}\\
  This annotation is based on the Linguistic Inquiry and Word Count -
  LIWC \cite{LIWC}, and it results to a classification of words along 64
  linguistic, cognitive, and affective categories .

\item {\bf Dialog act classification}\\
  With this annotation we classified the text into 15 dialog act classes
  that are based in the following taxonomy: Accept, Bye, Clarify,
  Continuer, Emotion, Emphasis, Greet, No Answer, Other, Reject,
  Statement, Wh-Question, Yes Answer, Yes/No Question, Order
  \cite{WACI}. Utterances that contained a url link, a empty utterances
  or utterances that did not include any ASCII characters were replaced
  by a "[url-link]" or "[empty-line]" tags, respectively.
\end{itemize}
{\bf Data availability}\\
The data are freely available for research purposes. They are provided as
supplementary material in a compressed "zip" file at
\url{http://www.sg.ethz.ch/downloads/Data}. If you have any
problems accessing them, please contact the authors.\\
{\bf Data structure, and the naming convention}\\
In the zip file each folder contains annotations of each one of the 20
IRC channels.  The file names correspond to the date, and their
extensions represent the type of annotation they provide. The general
internal structure of every file is as follows:
\begin{verbatim}
[timestamp] <annonymized-user-ID>  a file-type specific annotation
\end{verbatim}
More specifically, the type of information provided by every file is the following:\\
\newline
file extension: ``.sent"
\begin{verbatim}
[time-stamp] <userID> | sentimentClass | positiveArousal | negativeArousal |
[03:45]      <3032>   | 0              | 1               | -1              |
\end{verbatim}
file extension: ``.liwc"
\begin{verbatim}
[time-stamp] <userID> liwcCategory1:liwcCategory2:liwcCategory3
[03:45] <3032> Affect:Posemo:Assent
\end{verbatim}
file extension: ``.da"
\begin{verbatim}
[time-stamp] <userID> dialogActClass
[03:45] <3032> Emotion
\end{verbatim}

\newpage
\subsection*{Model details}

In order to understand how each one affects the ratio of emotion
polarities in the posts and the user and conversation persistence, we
performed a large set of simulations using different combination of
parameter values for the model described in Section "\emph{An agent-based
  model for chatroom users}". For each combination of values we performed
run 10 simulation sets, and the dependencies of the collective behavior
of the chatroom versus individual parameters are shown in Supplementary
Figures S\ref{Fig.S:M1}-S\ref{Fig.S:M3}.\\

In Supplementary Figure S\ref{Fig.S:M1} is summarized the effect in the
ratio of positive, negative and neutral posts due to the change in some
of the parameters. A higher amplitude of the stochastic influence implies
a lower frequency of neutral posts, splitting the rest equally among
positive and negative. Due to the high stochasticity of values like $A_v
= 0.4$, the community just behaves randomly with almost even ratios of
positive, negative and neutral. Increasing $b$, $c$, or decreasing the
decay of the field $\gamma_h$, the influence of the conversation in the
individual valence increases, leading to higher values of emotional posts
regardless of their polarity. An increase of the absolute value of the
expression thresholds $V_{\pm}$ yields a lower frequency of the
corresponding polarity, as expected.\\

In terms of valence decay, $\gamma_v$, we simulated two possible
cases. The case $\gamma_v = 0.1$ represents a virtual study of the
dynamics of mood, as a slower, conscious process that influences the
overall emotional state. The second case, $\gamma_v = 0.5$ results to a
faster decay more representative of the dynamics of core affect, or fast
emotional states. Supplementary Figure S\ref{Fig.S:M2} shows the
distributions of conversation and individual persistence for all the
simulations with the ranges of values for the rest of the parameters. We
find the case of $\gamma_v = 0.5$ closer to reality as observed in IRC
channels, where persistence are significant but not as strong as they
would be for the other case.\\

For each simulated case, we calculated the persistence of each individual
expression as well as the persistence of the whole conversation. We find
that increasing levels of amplitude of the stochastic component of the
valence leads to slightly higher average individual persistence, but
does not affect much the overall conversation persistence. Similar to the
case of the polarity fractions, larger values of $b$, $c$, or lower
values of $\gamma_h$ have the effect of increasing persistence, as the
coupling induced by the conversation is stronger. Similarly, higher
values of the thresholds lead to lower conversation persistence due to
the higher probability of neutral expression.\\

Given this behavior of the model from, we focused on a particular set of
values to simulate conversations similar to actual IRC chats. We used
10000 agents in a conversation lasting 45000 time units, and performed 10
realizations of the model using the following set of parameters:
\begin{displaymath}
  \begin{array}{rrrrrrr}       
    V_{-}=-0.15, & V_{+}=0.05, & \gamma_v=0.2, &  A_v=0.2, &
    b = 0.01, & c=0.05, & \gamma_h=0.9  
  \end{array}
\end{displaymath}
The results of an extensive set of simulations with these parameters are
shown, and discussed in Section "\emph{An agent-based model for chatroom
  users}" of the main text.

\newpage
\begin{figure}
 \centering
%Figure S9
% \includegraphics[scale=0.25]{S.I.FIGs/ModelSI1}
 \includegraphics[scale=0.25]{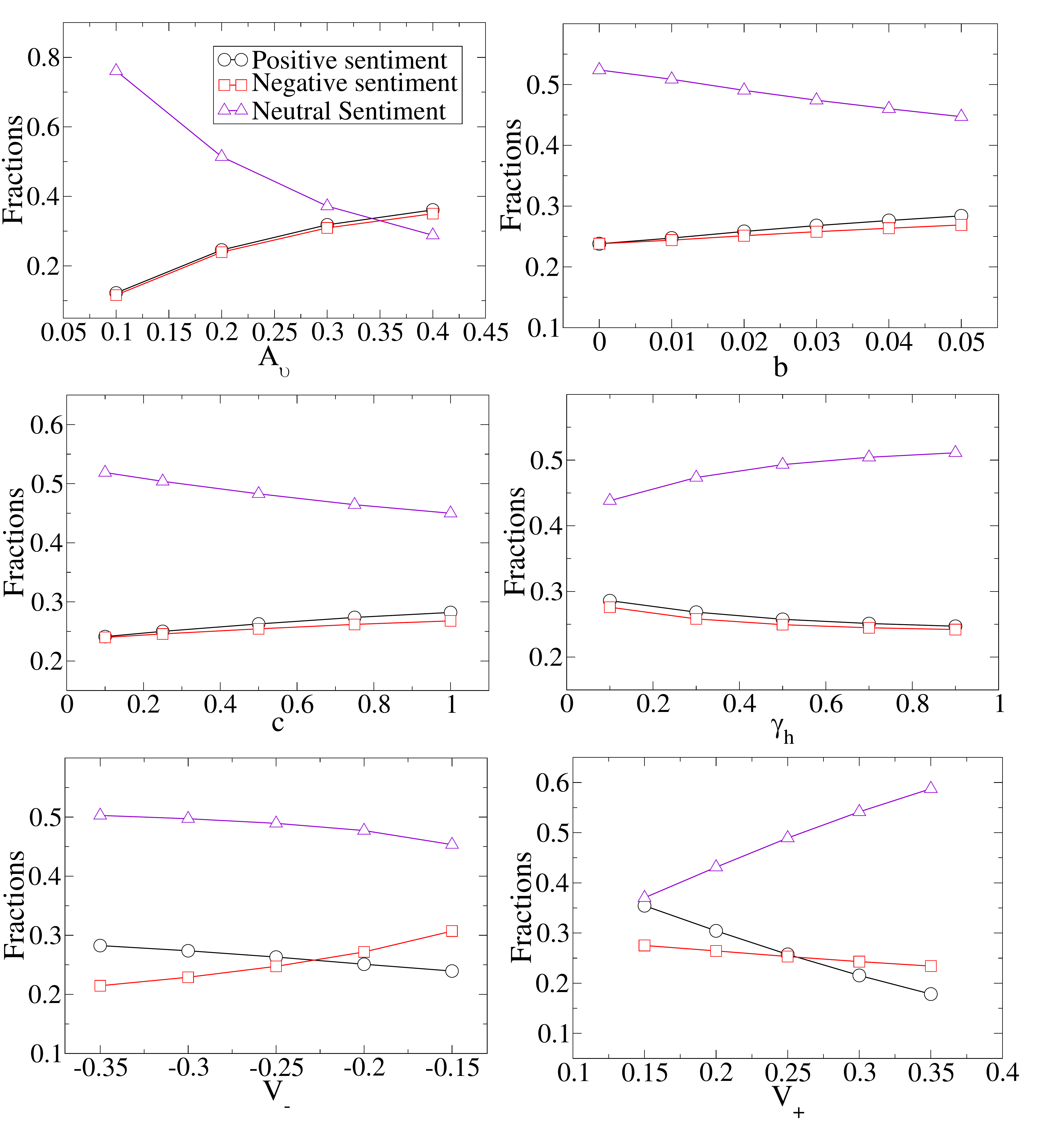}
 \caption{Fractions of positive, negative and neutral posts for different
   values of the parameters in our simulations. The ratio of emotional
   expressions (positive and negative) increases with the amplitude of
   the valence stochastic component $A_{v}$. This ratio is also slightly
   increased by the collective parameters $b$ and $c$, as the
   communication influence on the valence is stronger. The inverse is
   true for the parameter $\gamma_h$, i.e. the ratio of neutral posts
   increases the larger the decay of the field. An increase in the
   threshold $V_{\pm}$ leads to lower frequency of expression of the
   corresponding sign.}
 \label{Fig.S:M1}
\end{figure}

\begin{figure}
 \centering
%Figure S10
% \includegraphics[scale=0.55]{S.I.FIGs/ModelSI2}
 \includegraphics[scale=0.55]{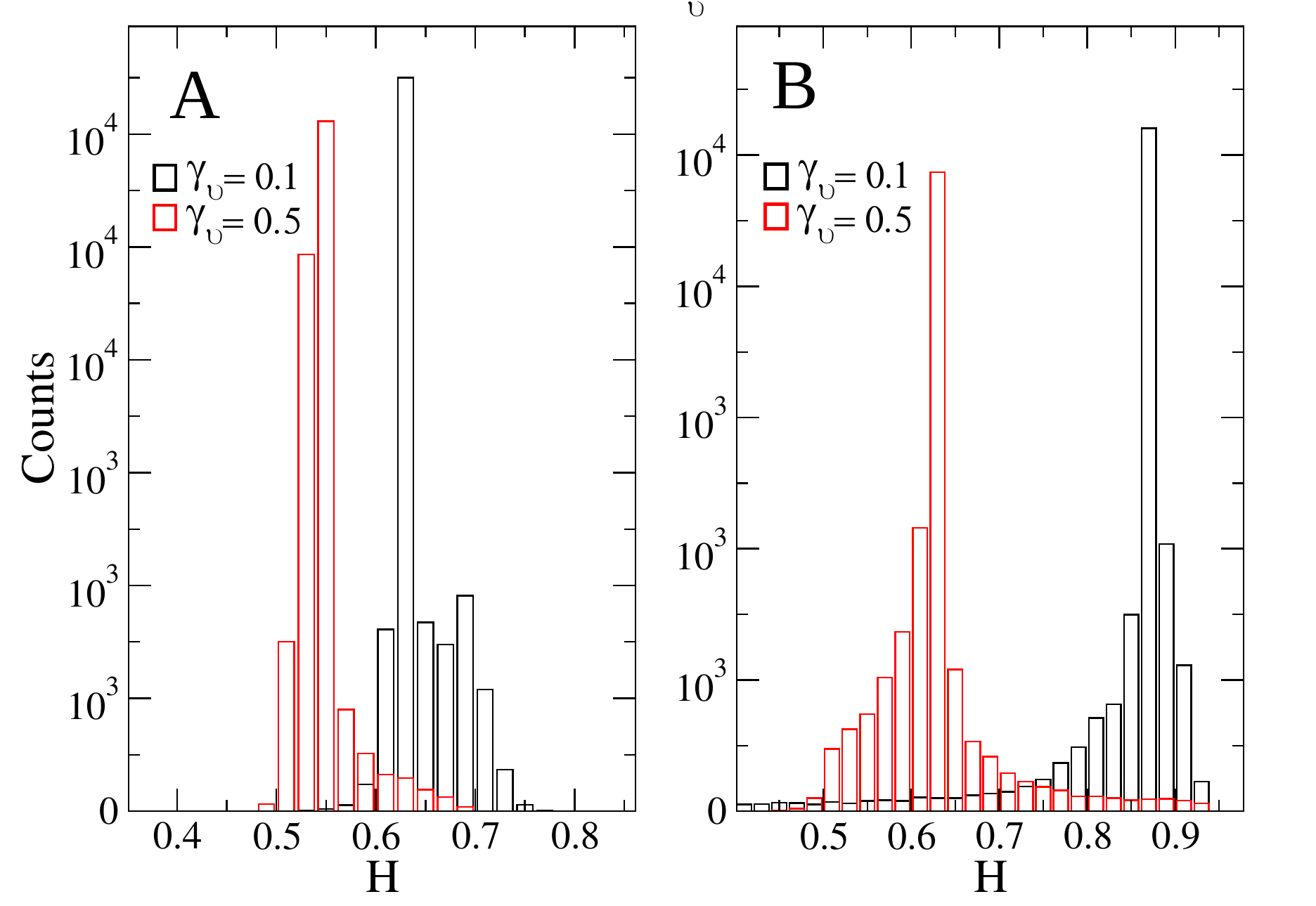}
 \caption{Distribution of the Hurst exponents for the simulated
   conversations (A) and agents (B) for the cases of $\gamma_v=0.1$ and
   $\gamma_v=0.5$. The Kolmogorov-Smirnov distance between the simulated
   distribution for $\gamma_v=0.1$ and the real data is KS=0.845, while
   between the KS distance between the simulated distribution for
   $\gamma_v=0.5$ and the real data is KS=0.519. This means that the
   individual and conversation persistence distributions are more similar
   to the real data (Fig 3 of the main text) for the case of
   $\gamma_v=0.5$, implying that the relaxation speed of the valence of
   chatroom users is fast.}
 \label{Fig.S:M2}
\end{figure}

\begin{figure}
 \centering
%Figure S11
% \includegraphics[scale=0.25]{S.I.FIGs/ModelSI3}
 \includegraphics[scale=0.25]{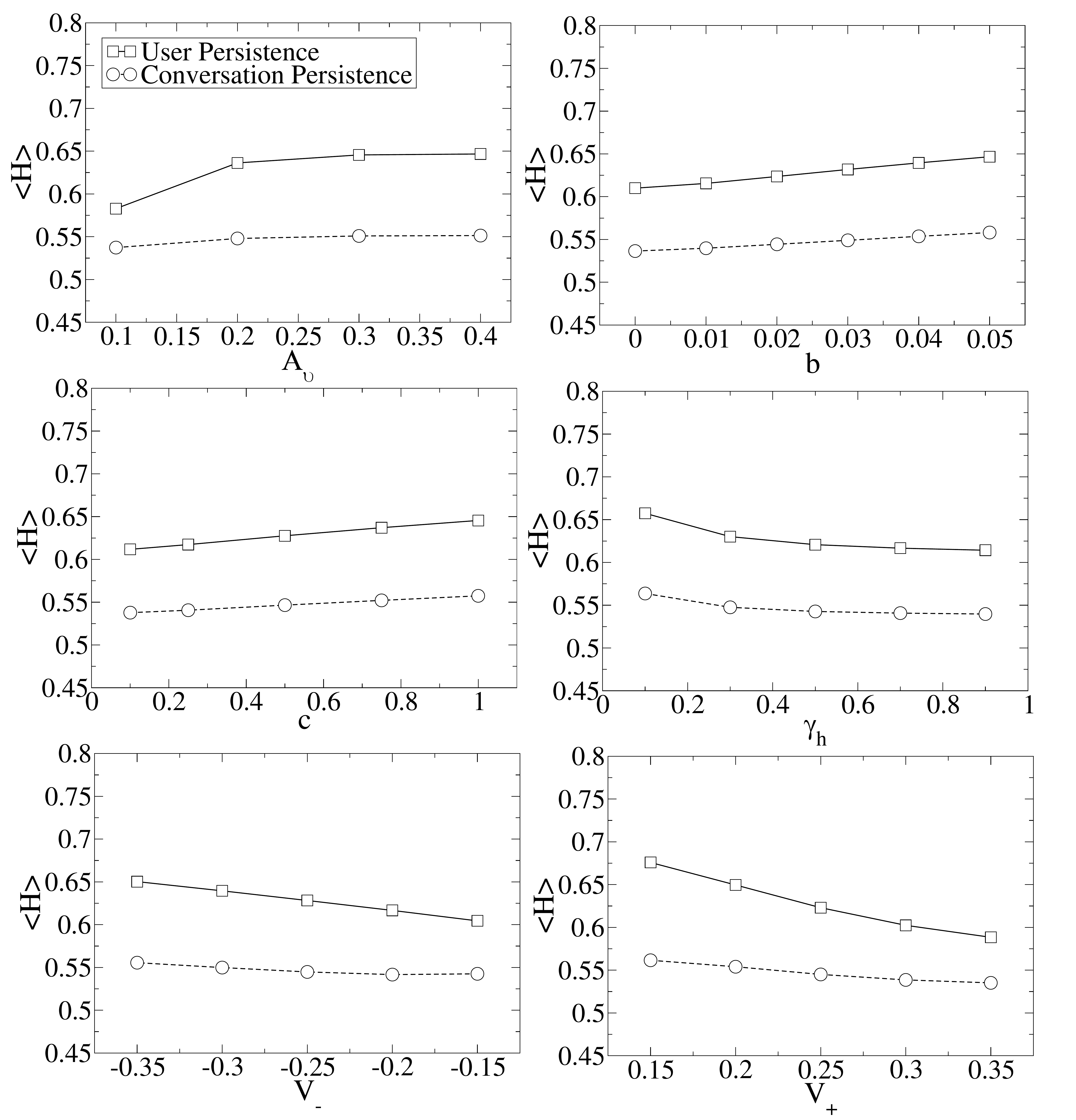}
 \caption{Mean value of the Hurst exponents of the emotional expression
   of agents and conversations for different values of the simulation
   parameters.  Under the influence of an emotional field, the user
   persistence increases with $A_v$, meaning that a stronger stochastic
   component can lead to conversations more similar to the observed
   ones. The coupling parameters $c$ and $b$ increase both mean
   persistence. The effect of larger $\gamma_h$ is the inverse, the
   stronger the decay of information, the weaker the persistence. Larger
   positive thresholds $V_{+}$ lead to lower user persistence, while the
   inverse is true for the negative threshold $V_{-}$. The standard error
   bars showing the standard error of the mean value are smaller that the
   symbol size and are not visible.}
 \label{Fig.S:M3}
\end{figure}

\begin{figure}
 \centering
%Figure S12
%Figure SXX2
 \includegraphics[scale=0.35]{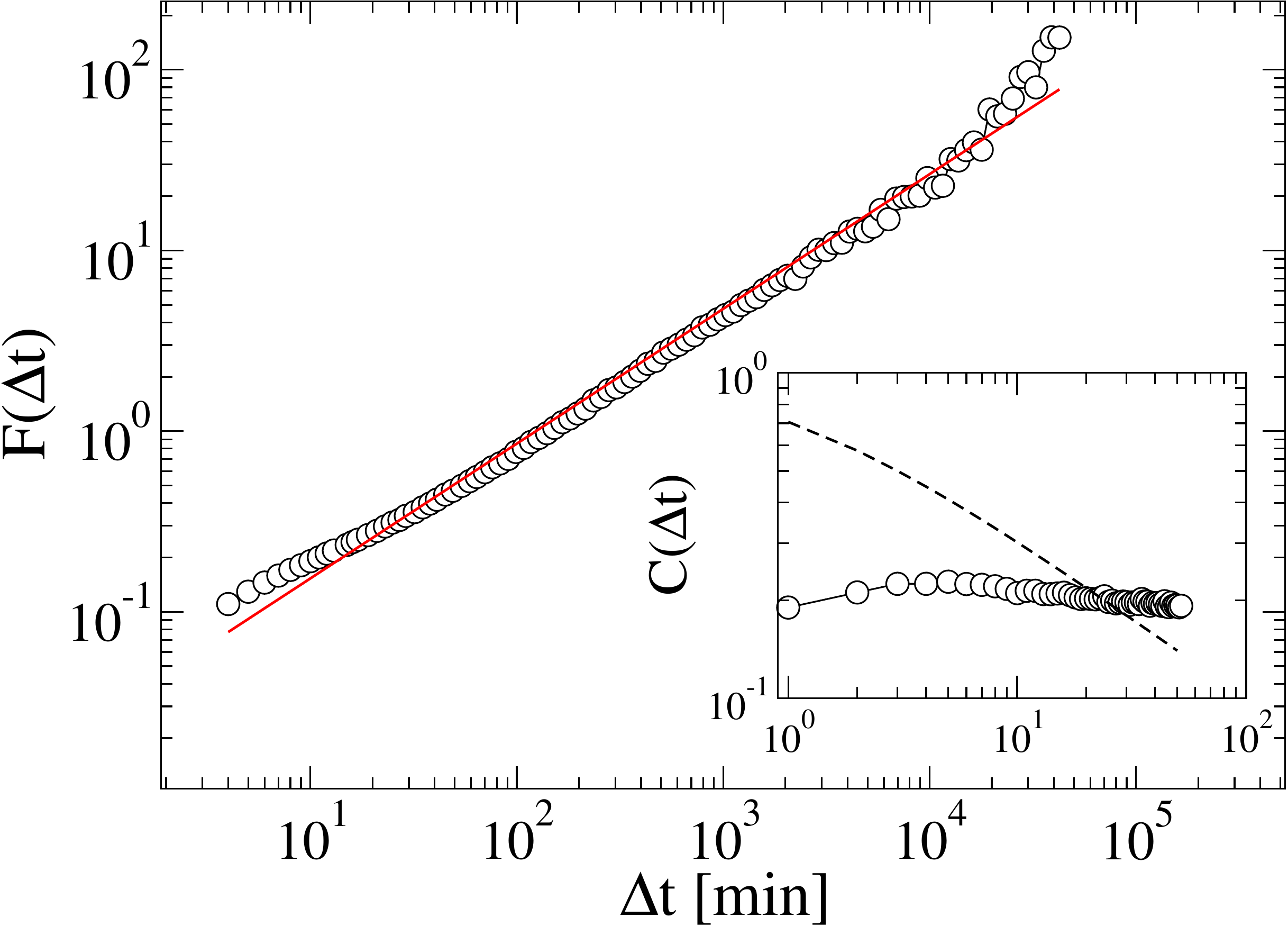}
 \caption{DFA fluctuation function calculated using the inter-event times
   of a simulated IRC channel. The Hurst exponent obtained is
   $H_{\omega'}\simeq 0.75$, suggesting the existence of log term
   correlations in the time series. We note the absence of pronounced
   dependencies in the user activity that would be manifested by a power
   law decaying autocorrelation function (Inset). The dotted line shows
   the expected decay of the autocorrelation function according to the
   scaling relation $\nu_{\omega'}=2-2H_{\omega'}$
   \cite{Kantelhardt2009}. In this case, the origin of the correlations
   revealed by the Hurst exponent can only be the broad distribution of
   inter-event times that was given as input to the model, since there is
   no coupling in the activity of users.}
 \label{Fig.S:XX2}
\end{figure}

\clearpage
\newpage

\bibliographystyle{pnas2009} %\bibliography{IRC-Channels-Results-PNAS} 

\end{document}